\documentclass[pre,aps,twocolumn,superscriptaddress,showpacs,floatfix,amssymb,amsmath]{revtex4-1}
\usepackage{epsfig}
\usepackage{graphics}
\usepackage{subcaption} %for subfigures
\usepackage[usenames,dvipsnames]{color}
\usepackage{caption}

\usepackage{float} %for fixing pictures in certan place in text
\usepackage{array}
\usepackage{multirow}

\usepackage{yfonts}

\usepackage{graphicx}
\usepackage{amssymb}
\usepackage{mathrsfs}
\usepackage{bbm}
\usepackage{epsfig}
\usepackage{makecell}

\usepackage[usenames,dvipsnames]{color}

\usepackage{bbm}
\usepackage{color}

\usepackage{kpfonts}
      
\definecolor{MyWhite}{RGB}{255,250,240} 
\definecolor{light-blue}{RGB}{240,248,255}
\definecolor{light-gray}{gray}{0.95}

\newcommand{\beq}{\begin{equation}}
\newcommand{\eeq}{\end{equation}}

\begin{document}

%\pagenumbering{empty}
%\begin{titlepage}
%
\title{%LARGE REVIEW VERSION \\ ~ \\
 \Large  Can all-atom protein dynamics be reconstructed \\
  from  the knowledge of C$\alpha$ time evolution?}

\vskip 5.0cm
\author{Jiaojiao Liu}
\email{ljjhappy1207@163.com}
\affiliation{School of Physics, Beijing Institute of Technology, Beijing 100081, P.R. China}
\author{Jin Dai}
\email{djcn1987@outlook.com}
\affiliation{Nordita, Stockholm University, Roslagstullsbacken 23, SE-106 91 Stockholm, Sweden}
\author{Jianfeng He}
\affiliation{School of Physics, Beijing Institute of Technology, Beijing 100081, P.R. China}
\email{hjf@bit.edu.cn}
\author{Xubiao Peng}
\email{xubiaopeng@gmail.com}
\affiliation{Center for Quantum Technology Research, School of Physics, 
Beijing Institute of Technology, Beijing 100081, P. R. China}
\author{Antti J. Niemi}
\email{Antti.Niemi@physics.uu.se}
\homepage{http://www.folding-protein.org}
\affiliation{Nordita, Stockholm University, Roslagstullsbacken 23, SE-106 91 Stockholm, Sweden}
\affiliation{Department of Physics and Astronomy, Uppsala University,
P.O. Box 516, S-75120, Uppsala, Sweden}
\affiliation{Laboratory of Physics of Living Matter, School of Biomedicine, Far Eastern Federal University, Vladivostok, Russia}
\affiliation{School of Physics, Beijing Institute of Technology, Beijing 100081, P.R. China}

\begin{abstract}
\noindent
We inquire to what extent  protein peptide plane and side chain dynamics can be reconstructed  from knowledge of 
C$\alpha$  dynamics. Due to lack of experimental data we analyze all atom molecular dynamics 
trajectories from {\it Anton} supercomputer,  and for clarity we limit our attention to the peptide plane O atoms and
side chain C$\beta$ atoms. We try and reconstruct their dynamics using four different approaches. Three of these are the
publicly available reconstruction programs {\it {\it Pulchra}},  {\it {\it Remo}}  and {\it {\it Scwrl4}}. 
The fourth, {\it Statistical Method},  builds entirely on statistical analysis of Protein Data Bank (PDB) structures.  
All four methods place the O and C$\beta$ atoms accurately along the {\it Anton} trajectories. However,
the {\it  Statistical Method} performs  best.  The results suggest that under physiological conditions,  the all atom dynamics  
is slaved to that of C$\alpha$ atoms. The results can help improve  all atom
force fields, and advance reconstruction and refinement methods for reduced protein structures. 
The results  provide impetus for development of effective coarse grained force fields  
in terms of reduced coordinates.
\end{abstract}

%\pacs{87.14.et, 02.40.-k, 02.90.+p, 05.65.+bk}

\maketitle

\section{Introduction}
%Some proteins function correctly once they become folded into a definite 
%conformational state. Other proteins remain highly dynamical when they exert their biological function. Some
%proteins even continue their movements after they become affixed into oligomers or bind ligands.  

%It should foster advances in biochemistry and pharmacology on a wide front, whenever
%the dynamical properties of a protein play a role.

The  structure of a protein is commonly characterized in terms of the C$\alpha$ atoms. They
are  located  at the branch points between the backbone and the side chains, and as such their positions are subject 
to relatively stringent stereochemical constraints. For example the model building in a crystallographic  structure determination 
experiment commonly starts with an initial C$\alpha$ skeletonization  \cite{Jones-1991}. The central role of the C$\alpha$ atoms 
is also exploited widely in various structural classification  schemes  \cite{Sillitoe-2013,Murzin-1995}, in threading \cite{Roy-2010}, 
homology \cite{Schwede-2003} and other modeling techniques \cite{Zhang-2009}, and  {\it de novo}  approaches \cite{Dill-2007}. The
development of coarse grained energy functions for folding prediction also frequently 
points out  the special role of C$\alpha$ atoms \cite{Scheraga-2007,Kmiecik-2016},
and the aim of the so-called C$\alpha$ trace problem is to construct an accurate 
main chain and/or all atom model of a crystallographic folded protein, solely  from the knowledge of the positions 
of the C$\alpha$ atoms \cite{Holm-1991,DePristo-2003,Lovell-2003,Rotkiewicz-2008,Li-2009,Krivov-2009}. 

In the dynamical case, knowledge of the all atom structure is pivotal to the understanding how biologically active proteins function. 
However, in the case of a dynamical protein it remains very hard to come by with high precision structural information, and as a consequence 
our understanding of protein 
dynamics remains very limited \cite{Henzler-2007,Frauenfelder-2009,Bu-2011,Khodadadi-2015}.  Here we test a widely suggested proposal that
the dynamics of backbone and side chain atoms could be strongly slaved to the dynamics of the C$\alpha$ atoms, under physiological conditions. For 
this we  address a {\it dynamical} variant of the static C$\alpha$-trace problem: We inquire to what extent can
the motions of the peptide plane and the side chain atoms be estimated from the knowledge of the C$\alpha$ atom positions, in a 
dynamical protein that moves under physiological conditions.  Any systematic correlation between the dynamics of C$\alpha$ atoms and other heavy atoms 
could be most valuable, for our understanding of many important biological processes.  A slaving of the peptide plane and side chain heavy atom motions to the  
C$\alpha$ backbone dynamics would mean that many aspects of protein dynamics can be described by effective coarse grained  energy 
functions that are  formulated in terms of reduced sets of coordinates that relate to  the C$\alpha$ atoms only.

Unfortunately, high precision experimental data on dynamical proteins under physiological conditions is sparse, indeed almost non-existent. 
At the moment all atom molecular dynamics simulations remain the primary source of dynamical  information. These simulations are 
best exemplified by the very long {\it Anton} trajectories \cite{Lindorff-2011}, that use the CHARMM22$^\star$ force field \cite{Piana-2011}.
 Accordingly we analyze the all atom  trajectories that were produced in these simulations; specifically we consider the $\alpha$-helical 
villin and the $\beta$-stranded ww-domain trajectories
reported in \cite{Lindorff-2011}. 
From the {\it Anton} trajectories, we extract the C$\alpha$ dynamics. We then try to reconstruct  the motions of 
other atoms: For clarity, we limit our attention to peptide plane O and the side chain C$\beta$ atoms only. We reckon this is a limitation, but these 
two atoms are common to all  amino acids except for glycine that lacks the C$\beta$ atom.  Moreover, the O atom does not share a covalent
bond, either with the C$\alpha$ or the C$\beta$.
In the {\it static} case,  the knowledge of  the C$\alpha$, O and C$\beta$ atoms is often considered sufficient to determine the positions of 
the remaining atoms, reliably and often at very high precision {\it e.g.}  with the help of stereochemical constraints  and
rotamer libraries \cite{Janin-1978,Lovell-2000,Schrauber-1993,Dunbrack-1993,Shapovalov-2011,Peng-2014}. Furthermore,
there are highly predictive coarse grained and associative memory Hamiltonians for protein structure determination that 
employ exactly the reduced coordinate set of $\mathrm C\alpha - \mathrm C\beta - \mathrm O$ atoms \cite{Sasai-1990,Davtyan-2012}.

We compare four different reconstruction methods.  These include the three  publicly available programs {\it Pulchra} \cite{Rotkiewicz-2008}
{\it Remo} \cite{Li-2009} and  {\it Scwrl} \cite{Krivov-2009}; note that {\it Scwrl} does not predict the peptide plane
atom positions, instead it is commonly used in combination with {\it Remo} to predict the side chain atom positions. 
The fourth approach we  introduce here. It is a pure {\it Statistical Method}, it  is  based entirely on information that we extract from high resolution 
crystallographic PDB structures. 

We find that all four methods have a very high success rate 
in predicting the positions of the dynamic O and C$\beta$ atoms along the {\it Anton} C$\alpha$ backbone, both in the case of villin and ww-domain. 
Surprisingly, we find that our straightforward {\it Statistical Method} performs even better than the three other much more elaborate 
methods. Since  our {\it Statistical Method} introduces no stereochemical fine tuning, nor force field
refinement, it  is also {\it superior}  to the other three in terms of computational speed.  

% We have found that the motions of the peptide plane and  side chain atoms are very  strongly correlated 
%with the C$\alpha$ atom positions,  even in a dynamical protein under physiological conditions; at least
%when the dynamics is described by all atom  {\it Anton} simulations \cite{Lindorff-2011}.

\vskip 0.2cm

\section{Methods}

\subsection{Anton data}

For data on protein dynamics, we use the all atom CHARMM22$^\star$ force field trajectories simulated with {\it Anton} supercomputer and 
reported in  \cite{Lindorff-2011}. The data has been provided to 
us by the authors.  We present detailed results for two trajectories that we have chosen for structural diversity:
We have selected the $\alpha$-helical villin (based on PDB structure 2F4K) and the $\beta$-stranded ww-domain
(based on PDB structure 2F21). The length of the villin trajectory is 120$\mu$s, and we have selected every 
20${th}$ simulated structure for our prediction analysis, for a total of 31395 structures. The length of the ww-domain trajectory is 651 $\mu$s and
we have chosen every 40${th}$ simulated structure for our prediction analysis, for a total of 60814 structures.
The combination of these two trajectories covers all the major regular secondary structures, with all the 
biologically relevant amino acids appearing, except CYS with its unique  potential to form  sulphur bridges. Furthermore,
the villin in \cite{Lindorff-2011} involves a NLE mutant and the HIS in \cite{Lindorff-2011} is protonated. Thus our analysis includes, at least
to some extent,  the effects of mutations and pH variations. In both villin and ww-domain, the {\it Anton} simulation observes several transitions
between structures that are unfolded and that are (apparently) folded.  This ensures that there is a good diversity of dynamical  
details for us to analyze. 

\subsection{Discrete Frenet frames}

\begin{figure}[h!]
\centering            
  \resizebox{6.5cm}{!}{\includegraphics[]{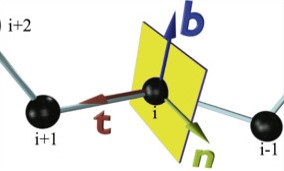}}
\caption {\small  {\it Color online:} 
Definition of Frenet frames (\ref{t})-(\ref{n}) at the position of a C$\alpha$ atom.
}   
\label{fig-1}
\end{figure}

Our basic tool of analysis is the discrete Frenet framing of the C$\alpha$ backbone \cite{Hu -2011} that we construct as follows: 
We take $\mathbf r_i$ ($i=1,...,N$) to be the (time dependent) coordinate of the $i^{th}$ C$\alpha$ atom along the backbone.
The $\mathbf r_i$ then form the vertices of the virtual C$\alpha$ backbone, that we visualize as a piecewise linear polygonal chain. 
At each vertex we introduce a right-handed orthonormal set of Frenet frames ($\mathbf n \mathbf b \mathbf t$)  that we
define as follows
\begin{equation}
\mathbf t_i = \frac{\mathbf r_{i+1} - \mathbf r_i}{|\mathbf r_{i+1} - \mathbf r_i |}
\label{t}
\end{equation}
\begin{equation}
\mathbf b_i = 
\frac{\mathbf t_{i-1} \times \mathbf t_i}{\mathbf t_{i-1} \times \mathbf t_i}
\label{b}
\end{equation}
\begin{equation}
\mathbf n_i = \mathbf b_i  \times \mathbf t_i
\label{n}
\end{equation}
The framing is shown in Figure \ref{fig-1}. For details of discrete Frenet frames and other 
framings of the C$\alpha$ backbone, we refer to \cite{Hu -2011}, and for a comparison with the 
Ramachandran angle description we refer to \cite{Hinsen-2013}.
%
%
%
%
%
%
%
%
%
%%%%%%%%%%%%%%%%%%%%%%%%%%%%%%%%%%%%%%%%%%%%%
%%
%%
%%
%%
%%
%%                           FIGURE 1
%%
%%
%%
%%
%%
%% 
%%%%%%%%%%%%%%
%
%
%
%
%
%%%%%%%%%%%%%%
%
%
%
%%
%{
%\footnotesize
%\begin{figure}[h]         
%\centering            
%  \resizebox{8 cm}{!}{\includegraphics[]{figure-1.jpg}}
%\caption {\small  {\it Color online:} 
%The three vectors ($\mathbf n, \mathbf b, \mathbf t$) form a righ-handed orthonormal coordinate system at the
%location $\mathbf r_i$ of each C$\alpha$ atom.
%%Definition of bond ($\kappa_i$) and torsion ($\tau_i$) angles in relation to the $i^{th}$ C$\alpha$  atom.
%}   
%\label{fig-1}    
%\end{figure}
%
%
%
%
%%%%%%%%%%
%%%%%%%%%%%%%%
%%%%%%%%%%%%%%%
%%%%%%%%%%%%%%
%%%%%%%%%%%%%%
%%%%%%%%%%%%%%%

\subsection{{\it Pulchra}, {\it Remo} and {\it Scwrl4} frames}

The framing used in {\it Pulchra} \cite{Rotkiewicz-2008}
is obtained from four successive C$\alpha$ coordinates. We start by defining
\begin{equation}
{\mathbf e}_{i,j} \ = \ \frac{ \mathbf r_i - \mathbf r_j}{ |  \mathbf r_i - \mathbf r_j|}
\label{eij}
\end{equation}
The {\it Pulchra} frames  ($\mathbf \nu_x^i, \nu_y^i, \nu_z^i$) are then
\begin{equation}
{\mathbf \nu}^i_x \ = \  {\mathbf e}_{i-1,i+1} 
\label{e1}
\end{equation}
\begin{equation}
{\mathbf \nu}^i_y \ = \ \frac{ \mathbf e_{i,i+1} \times \mathbf e_{i-1,i} }{|
\mathbf e_{i,i+1} \times \mathbf e_{i-1,i} |}
\label{e2}
\end{equation}
\begin{equation}
\mathbf \nu^i_z = \mathbf \nu^i_x \times \mathbf \nu^i_y
\label{e3}
\end{equation}
The {\it Remo} \cite{Li-2009} reconstruction program uses also frames that are obtained from four successive C$\alpha$ coordinates: We again
start with (\ref{eij}) and we set 

The {\it Remo} frames are then
\begin{equation}
{\mathbf x}_i   \ = \ \frac{ \mathbf e_{i-1,i+2} + \mathbf e_{i,i+1} }{ |  \mathbf e_{i-1,i+2} + \mathbf e_{i,i+1} |}
\label{u}
\end{equation}
\begin{equation}
\mathbf y_i \ = \ \frac{ \mathbf e_{i-1,i+2} - \mathbf e_{i,i+1} }{ |  \mathbf e_{i-1,i+2} - \mathbf e_{i,i+1} |}
\label{v}
\end{equation}
\begin{equation}
\mathbf z_i \ = \ \mathbf x_i \times \mathbf y_i 
\label{w}
\end{equation}
Both {\it Pulchra} and {\it Remo} frames  are very different from the discrete Frenet frames. In particular, both of these framings exploit 
four consecutive  C$\alpha$ atoms, while the discrete Frenet frames use only three.

Finally, in the case of {\it Scwrl4} the backbone is described in terms of the Ramachandran
angles \cite{Krivov-2009}.
   
%
%
%
%
%
%
%
%
%
%%%%%%%%%%%%%%%%%%%%%%%%%%%%%%%%%%%%%%%%%%%%%
%%
%%
%%
%%
%%
%%                           FIGURE 2
%%
%%
%%
%%
%%
%% 
%%%%%%%%%%%%%%
%%%%%%%%%%%%%%
%
%
%
%%
%{
%\footnotesize
\begin{figure}[h!]         
\centering            
  \resizebox{6.5cm}{!}{\includegraphics[]{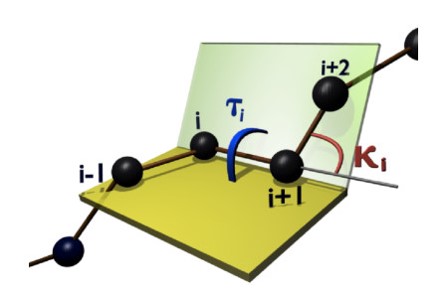}}
\caption {\small  {\it Color online:} 
Definition of bond ($\kappa_i$) and torsion ($\tau_i$) angles in relation to the $i^{th}$ C$\alpha$  atom.
}   
\label{fig-2}    
\end{figure}
%
%
%
%
%%%%%%%%%%
%%%%%%%%%%%%%%
%%%%%%%%%%%%%%%
%%%%%%%%%%%%%%
%%%%%%%%%%%%%%
%%%%%%%%%%%%%%%

\subsection{Visualization}

For protein structure visualization and our data analysis, we use exclusively the Frenet frames ($\mathbf n, \mathbf b, \mathbf t$); for 
each C$\alpha$ atom we associate a (virtual) C$\alpha$ backbone bond ($\kappa$) and torsion ($\tau$) angle as follows,
\begin{equation}
\begin{matrix}\vspace{0.2cm}
\kappa_i =  \arccos (\mathbf t_{i+1} \cdot \mathbf t_i)
\\
\tau_i =  {\rm sign} ( \mathbf b_{i+1} \cdot \mathbf n_i) \arccos (\mathbf b_{i+1} \cdot \mathbf b_i)
\end{matrix}
\label{kt}
\end{equation}
These angles are shown in Figure \ref{fig-2}. To  visualize the various atoms in a protein, we identify 
the bond and torsion angles ($\kappa,\tau$) as the canonical latitude 
and longitude angles on the surface of a unit radius (Frenet) sphere  $\mathbb S^2_\alpha$: The center of the sphere is 
located at the C$\alpha$ atom, the north-pole of  $\mathbb S^2_\alpha$  is the point where 
the latitude angle $\kappa=0$, the vector $\mathbf t$ points to the direction of the north-pole and this direction coincides with the
canonical $z$-direction in a C$\alpha$ centered Cartesian coordinate system. The great circle 
$ \tau=0$ passes through the north-pole and the  tip of the normal vector $\mathbf n$ that lies at the equator,   
the longitude {\it a.k.a.} torsion angle takes values $\tau\in [-\pi,\pi)$ and it increases in the counterclockwise direction 
around the positive $z$-axis {\it i.e.} around 
vector $\mathbf t$.

In the sequel, whenever we introduce a Frenet sphere $\mathbb S^2$ 
we shall use the convention that the triplet ($\mathbf n, \mathbf b, \mathbf t$)  corresponds to the right-handed Cartesian 
($xyz$)$\sim$($rgb$) coordinate system, with the convention that $\mathbf n \sim x \sim $ red ($r$), $\mathbf b \sim  y \sim $ green ($g$)
and $\mathbf t \sim z \sim $ blue ($b$). The color coding of distributions on the spheres are always relative but with equal {\it MatLab} setting, 
in all the cases that 
we display, and intensity increases from no-entry white to low density blue, and towards high density red.

We plot the directions of the peptide plane O and side chain C$\beta$ atoms  
on the surface of $\mathbb S^2_\alpha$ in {\it exactly}  the way how they are seen from the position 
of a miniature observer, standing at the center of the sphere $\mathbb S^2_\alpha$ and with head up towards the
north-pole.
For the O atoms we use  spherical coordinates that we denote ($\theta,\phi$) for the latitude and longitude, for the C$\beta$ atoms we denote the spherical
coordinates ($\vartheta,\varphi$) for the latitude and longitude on  $\mathbb S^2_\alpha$. Both sets of coordinates are in direct 
correspondence with ($\kappa,\tau$).

\begin{figure}[h!]         
\centering            
  \resizebox{6.5cm}{!}{\includegraphics[]{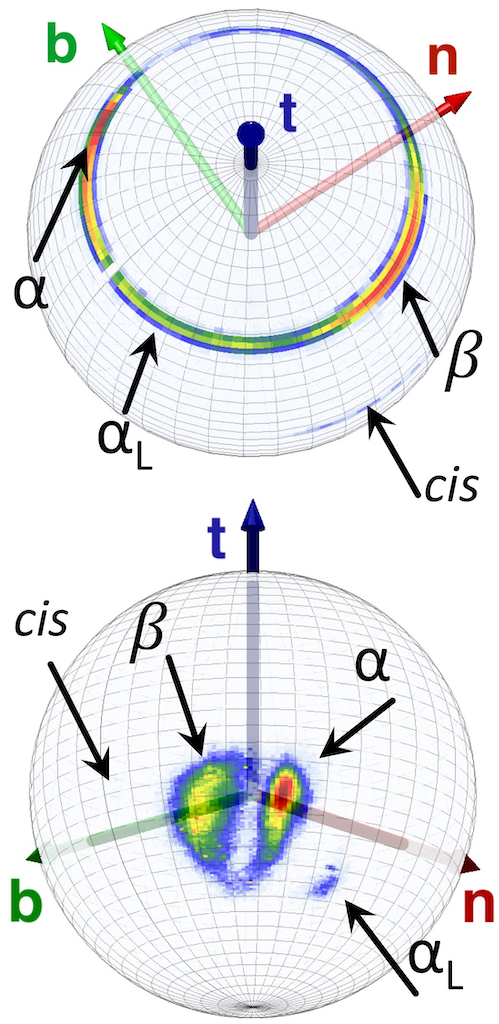}}
\caption {\small  {\it Color online:} 
Top: The ($\theta,\phi$) distribution of peptide plane O atoms in the below 1.0 \AA ~ resolution PDB structures on the surface of the Frenet sphere $\mathbb S^2_\alpha$. 
Bottom: The ($\vartheta,\varphi$) distribution of C$\beta$ atoms in  the below 1.0 \AA ~  resolution PDB structures on  $\mathbb S^2_\alpha$.  In both Fgures we have identified 
the major regions $\alpha$-helices ($\alpha$), $\beta$-strands ($\beta$), left-handed $\alpha$-helices ($\alpha_L$) and
{\it cis}-peptide planes ($cis$).
}   
\label{fig-3}    
\end{figure}

In Figure \ref{fig-3} we visualize the statistical reference distributions of the O and C$\beta$  atoms, in
crystallographic Protein Data Bank structures that have been 
measured with better than 1.0 \AA~ resolution. We choose these,  since we 
trust that such ultra high resolution structures are relatively 
void of refinement. 

%
%
%
%
%
%
%
%
%
%%%%%%%%%%%%%%%%%%%%%%%%%%%%%%%%%%%%%%%%%%%%%
%%
%%
%%
%%
%%
%%                           FIGURE  3
%%
%%
%%
%%
%%
%% 
%%%%%%%%%%%%%%
%%%%%%%%%%%%%%
%
%
%
%%
%{
%\footnotesize
%
%
%
%
%%%%%%%%%%
%%%%%%%%%%%%%%
%%%%%%%%%%%%%%%
%%%%%%%%%%%%%%
%%%%%%%%%%%%%%
%%%%%%%%%%%%%%%

The O distribution is concentrated on a very narrow circle-like  annulus. It forms the base of a right circular cone with axis that 
coincides with the $\mathbf t$ vector and with conical apex at the center of the Frenet sphere; the latitude of the annulus is very close to $\theta = \pi/4$ (rad)
so that the apex of the cone  is very close to $\pi/2$.
% \marginpar{{\textcolor{red}{\tiny Check numerical values}}}
The regions of $\alpha$-helical and $\beta$-stranded regular structures are connected by a region of left-handed $\alpha$-helical structures and by a region of
loops. 
There are very few entries outside this circular region; notably  the {\it cis}-peptide plane region is located on a short strip under the  $\beta$-stranded region
with a latitude angle close to $\theta \approx \pi/2$. 

The C$\beta$ distribution shown in Figure \ref{fig-3}  is a little like a  horse-shoe, forming a slightly distorted  annulus that is somewhat wider  than in the case of the O distribution. 
 The regions of $\alpha$-helical and $\beta$-stranded
regular structures are connected by a region of loops but the distribution of left-handed $\alpha$-helical  structures is now disjoint from the main distribution.

%
%
%
%
%
%
%
%
%
%%%%%%%%%%%%%%%%%%%%%%%%%%%%%%%%%%%%%%%%%%%%%
%%
%%
%%
%%
%%
%%                           FIGURE  4
%%
%%
%%
%%
%%
%% 
%%%%%%%%%%%%%%
%%%%%%%%%%%%%%
%
%
%
%%
%{
%\footnotesize
\begin{figure}[h!]         
\centering            
  \resizebox{6.5cm}{!}{\includegraphics[]{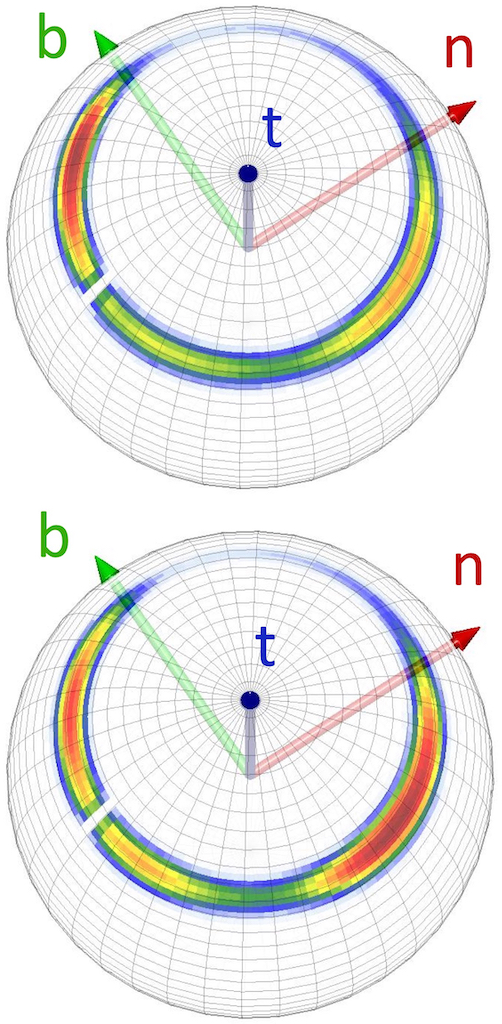}}
\caption {\small  {\it Color online:} 
Top: The distribution of peptide plane O atoms in the villin trajectory of {\it Anton}.  
Bottom: The distribution of peptide plane O atoms in the ww-domain  trajectory of {\it Anton}.
}   
\label{fig-4}    
\end{figure}
%
%
%
%
%%%%%%%%%%
%%%%%%%%%%%%%%
%%%%%%%%%%%%%%%
%%%%%%%%%%%%%%
%%%%%%%%%%%%%%
%%%%%%%%%%%%%%%

%

\begin{figure}[h!]         
\centering            
  \resizebox{6.5cm}{!}{\includegraphics[]{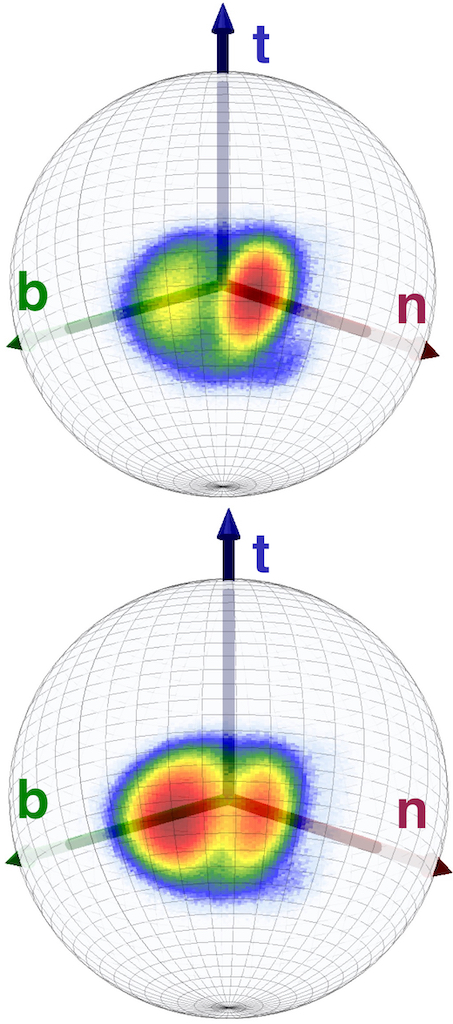}}
\caption {\small  {\it Color online:} 
Top: The distribution of side chain C$\beta$  atoms in the villin trajectory of {\it Anton}.  
Bottom: The distribution of side chain C$\beta$ atoms in the ww-domain  trajectory of {\it Anton}.
}   
\label{fig-5}    
\end{figure}

In Figures \ref{fig-4} and \ref{fig-5} we show the corresponding  {\it Anton} distributions for the O an C$\beta$ atoms, separately in the case of villin and ww-domain.
When we compare the {\it Anton} distributions and the ultra high resolution PDB data of Figure \ref{fig-3}, we observe 
that the overall structure of the statistical distributions are very similar. We also note that as expected, in the villin trajectory there is a clear predominance of
the $\alpha$-helical region, while in the case of ww-domain the $\beta$-stranded region dominates. The PDB and {\it Anton} distributions are otherwise remarkably similar,
superficially the only difference appears to be due to thermal fluctuations in the latter:
The crystallographic data is  often taken at liquid nitrogen temperatures below 77 K  while the simulation temperature in the {\it Anton} data is around 360K for both 
villin and ww-domain.

\noindent
The strong similarity between the distributions in Figures \ref{fig-3}-\ref{fig-5} motivates us to make the following bold proposal:  Even in the case of a dynamical protein
under near-physiological conditions, the relative positions of the O and C$\beta$ atoms can be reconstructed with high accuracy from the knowledge of the C$\alpha$ 
atoms motion  - at least when the CHARMM22$^\star$ force field is used to describe the dynamics.

%%%%%%%%%%%%%%
%%%%%%%%%%%%%%
%%%%%%%%%%%%%%%
%
%
%
%
%
%
%
%
%
%%%%%%%%%%%%%%%%%%%%%%%%%%%%%%%%%%%%%%%%%%%%%
%%
%%
%%
%%
%%
%%                           FIGURE  5
%%
%%
%%
%%
%%
%% 
%%%%%%%%%%%%%%
%%%%%%%%%%%%%%
%
%
%
%%
%{
%\footnotesize
%
%  %
%
%%%%%%%%%%
%%%%%%%%%%%%%%
%%%%%%%%%%%%%%%
%%%%%%%%%%%%%%
%%%%%%%%%%%%%%
%%%%%%%%%%%%%%%
%
%
%
%
%
%
%
%
%

\subsection{Reconstruction}

We proceed to try and reproduce the individual O and C$\beta$ atom positions in the {\it Anton} trajectories of villin and ww-domain
\cite{Lindorff-2011}, solely from the knowledge of the C$\alpha$ atoms.  We employ four different reconstruction methods:

\subsubsection{Pulchra}  

For {\it Pulchra} \cite{Rotkiewicz-2008}  we use the version 3.04. 
We start with the C$\alpha$ coordinates that we obtain from {\it Anton}. We then use {\it Pulchra}  to reconstruct the other heavy atom
positions, with only the instantaneous C$\alpha$ coordinates of the {\it Anton}  trajectory as an  input.  From the {\it Pulchra} structures
we read the coordinates of the peptide plane O and side chain C$\beta$  atoms, and
compare with the original {\it Anton} data.

 \subsubsection{{\it Remo}}
  
 For {\it Remo} \cite{Li-2009} we use the version 3.0, and we proceed  with reconstruction as in the case of  {\it Pulchra}. 
 We note that  {\it Remo} employs {\it Scwrl} for the side chains.

 \subsubsection{{\it Scwrl4}} 
  
 For stand-alone {\it Scwrl} \cite{Krivov-2009} we use the version 4.0.  Since
 {\it Scwrl} can not reconstruct the peptide planes, we use it only for the C$\beta$ comparison. For this we first construct the peptide
  planes  using  {\it Pulchra}, since {\it Remo} is already based on {\it Scwrl}.

  \subsubsection{Statistical Method}
  
Unlike the previous three methods, our {\it Statistical Method } approach to reconstruction 
does not employ any force field refinement, stereochemical constraints,  or any other kind of data curation. It only
uses statistical analysis of PDB data to predict the positions of the  peptide 
plane O and side chain C$\beta$ atoms  from the knowledge of the C$\alpha$ atom positions: Once the C$\alpha$ coordinates
of an amino acid are given,  a search algorithm fits it with a PDB structure and identifies the ensuing  
O and C$\beta$ atom coordinates as the reconstructed coordinates.  

As already stated, the PDB pool consist of all those crystallographic structures that 
have been measured with better than 1.0 \AA~ resolution. 
We start with a visual presentation of the  C$\alpha$ bond and torsion 
angle density distribution of these PDB structures, in terms of a stereographically projected Frenet sphere. 

Let 
$\mathbb S^2_\alpha(i)$ be  centered at the $i^{th}$ C$\alpha$ atom of a given
PDB structure. The vector $\mathbf t_i$ has its tail at the center of $\mathbb S^2_\alpha(i)$ and its head lies at the north pole,
this vector points from the $i^{th}$ C$\alpha$ atom towards  the $(i+1)^{st}$ C$\alpha$ atom.  
The $(i+2)^{nd}$ C$\alpha$ is then located similarly,  in the direction of the 
Frenet vector $\mathbf t_{i+1}$ that points  from the center of $\mathbb S^2_\alpha(i+1)$ towards its north pole.  

We parallel transport the vector $\mathbf t_{i+1}$ without any rotation until its tail becomes located at the $i^{th}$ C$\alpha$ atom
position {\it i.e.}  at the origin of $\mathbb S^2_\alpha(i)$.  Let ($\kappa_i,\tau_i$) be the Frenet frame coordinates of the head of the parallel transported  $\mathbf t_{i+1}$  
on the surface of $\mathbb S^2_\alpha(i)$. These coordinates depict how a miniature  Frenet frame
observer, standing at  the position of the $i^{th}$ C$\alpha$ atom and head towards the north pole of $\mathbb S^2_\alpha(i)$,  sees the backbone twisting and bending when she 
proceeds along the chain to the position of the $(i+1)^{st}$ C$\alpha$ atom.

\begin{figure}[h!]         
\centering            
  \resizebox{6.5cm}{!}{\includegraphics[]{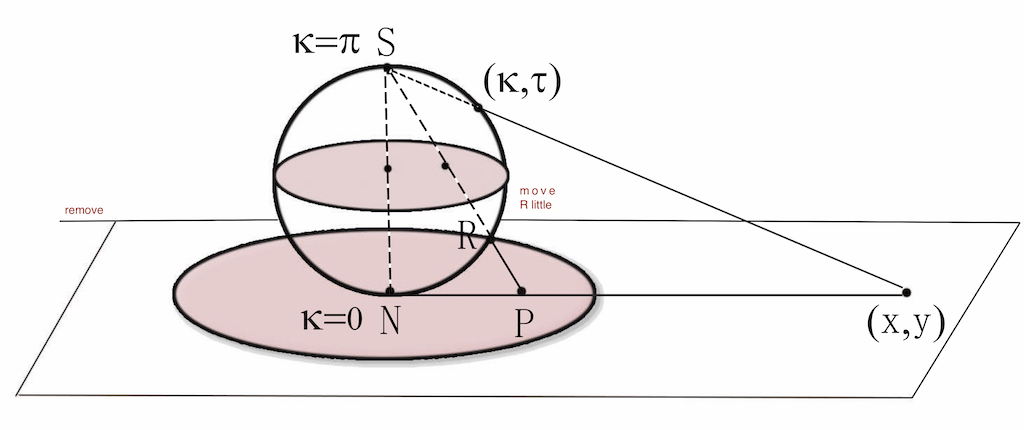}}
\caption {\small  {\it Color online:} 
Stereographic projection of two sphere onto plane, with the projection taken from the south-pole. 
}   
\label{fig-6}    
\end{figure}

We repeat the construction for all the C$\alpha$ atoms along all the chains in our pool.   
This yields us a statistical distribution   
in terms of the coordinates ($\kappa,\tau$)  of the heads of the  parallel transported vectors $\mathbf t$.  We 
visualize the distribution by projecting the 
sphere $\mathbb S^2_\alpha$ stereographically onto the complex plane 
from the south pole  as shown in Figure \ref{fig-6}.  
%%%%%%%%%%%%%%
%%%%%%%%%%%%%%%
%
%
%
%
%
%
%
%
%
%%%%%%%%%%%%%%%%%%%%%%%%%%%%%%%%%%%%%%%%%%%%%
%%
%%
%%
%%
%%
%%                           FIGURE  6
%%
%%
%%
%%
%%
%% 
%%%%%%%%%%%%%%
%%%%%%%%%%%%%%
%
%
%
%%
%{
%\footnotesize
%
%
%
%
%%%%%%%%%%
%%%%%%%%%%%%%%
%%%%%%%%%%%%%%%
The relation
between the spherical coordinates ($\kappa,\tau$) and the $z=x+iy$ coordinates on the plane  is
\[
x+iy = \tan\left( \frac{\kappa}{2}\right) e^{i\tau}
\]

\begin{figure}[h!]         
\centering            
  \resizebox{6.5cm}{!}{\includegraphics[]{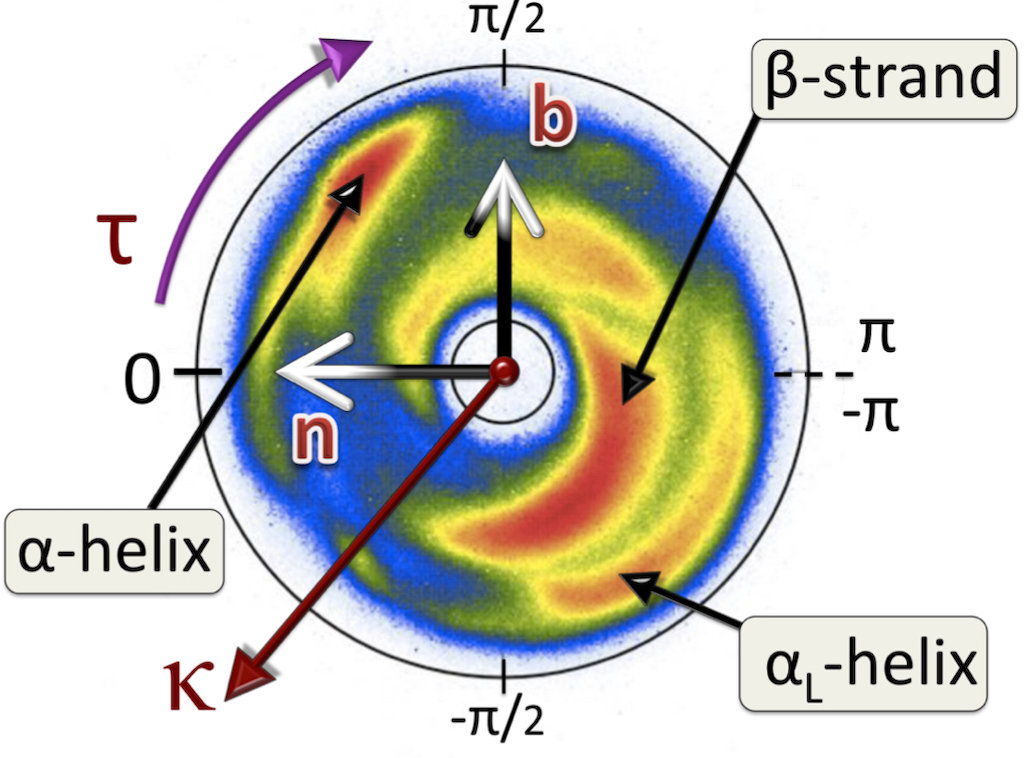}}
\caption {\small  {\it Color online:} 
Statistical distribution of all C$\alpha$ atoms in our 1.0 \AA~ pool of PDB structures, on stereographically projected 
sphere shown in Figure \ref{fig-6}. 
}   
\label{fig-7}    
\end{figure}

In Figure \ref{fig-7} we show the  distribution of all the C$\alpha$ atom coordinates in our PDB data set, on the stereographically projected
Frenet sphere.
%%%%%%%%%%%%%%
%%%%%%%%%%%%%%%
%%%%%%%%%%%%%%%%%%%%%%%%%%%%%%%%%%%%%%%%%%%%%
%%
%%
%%
%%
%%
%%                           FIGURE  7
%%
%%
%%
%%
%%
%% 
%%%%%%%%%%%%%%
%%%%%%%%%%%%%%
%
%
%
%%
%{
%\footnotesize
%
%
%
%
%%%%%%%%%%
%%%%%%%%%%%%%%
%%%%%%%%%%%%%%%
The distribution is largely concentrated inside an annulus, with inner circle  $\kappa_{in} \approx 1$ and outer circle
$\kappa_{out} \approx \pi/2$. The various regular secondary structures such as $\alpha$-helices, 
$\beta$-strands and left-handed $\alpha$-helices are clearly identifiable  and marked in this Figure. 

We proceed to describe how we predict the positions of the  O and C$\beta$ atoms from the knowledge of the  C$\alpha$ atoms coordinates,
along a given (dynamical) protein structure: We first use the torsion angle $\tau \in[-\pi, \pi) $ to divide
the  statistical distribution of Figure \ref{fig-7}  into 60 equal size sectors sized $\Delta\tau = \pi/30$ radians.  
We then divide each of these sectors into two sets, one with  bond angle  $\kappa < 1.2$ (rad) 
and the other with $\kappa \geq 1.2$ (rad); we choose this value since the circle $\kappa = 1.2$ divides the annulus in 
Figure \ref{fig-7} roughly into annuli of $\alpha$-helix-like and
$\beta$-strand-like (secondary) structures.  

Now suppose that we have a C$\alpha$ atom along a protein backbone, 
with coordinates ($\kappa_i, \tau_i$).  We determine the coordinates
($\theta_i,\phi_i$) of the corresponding O atom and the coordinates  ($\vartheta_i,\varphi_i$) of the corresponding  C$\beta$ atom using the following
algorithm:

\vskip 0.2cm
\noindent
{\bf Step 1:~}   We first use the $\tau_i$ value of the C$\alpha$ atom 
to select the pertinent sector $\tau_i \in \Delta \tau$. We then use its  $\kappa_i$ value  together with 
$\kappa_{i+1}$ of the following C$\alpha$ atom along the chain, to assign with the given C$\alpha$ atom one of the four sets
\[
\begin{matrix} 
{\rm Set ~ \Delta\kappa_1:} ~~~~~~~~ \kappa_{i} \ < \ 1.2 \ \ \ \ \& \ \ \ \  \kappa_{i+1} \ < \ 1.2 \\
{\rm Set ~ \Delta\kappa_2:} ~~~~~~~~ \kappa_{i} \ < \ 1.2 \ \ \ \ \& \ \ \ \  \kappa_{i+1} \ \geq \ 1.2 \\
{\rm Set ~ \Delta\kappa_3:} ~~~~~~~~ \kappa_{i} \ \geq \ 1.2 \ \ \ \ \& \ \ \ \  \kappa_{i+1} \ < \ 1.2 \\
{\rm Set ~ \Delta\kappa_4:} ~~~~~~~~ \kappa_{i} \ \geq \ 1.2 \ \ \ \ \& \ \ \ \  \kappa_{i+1} \ \geq \ 1.2
\end{matrix}
\]
Together with the $\Delta \tau$ sectors this divides our statistical C$\alpha$ distribution
into 4$\times$60 subsets $[\Delta\kappa;\Delta\tau]$, and we choose the one that corresponds 
to the  ($\kappa_i, \kappa_{i+1}, \tau_i$) values of the C$\alpha$ atom we  consider.

\vskip 0.2cm
\noindent
{\bf Step 2:~}    We search a protein structure in our pool, in the subset $[\Delta\kappa,\Delta\tau]$ of the $i^{th}$ C$\alpha$, for which two consecutive amino acids 
are also identical  to the $i^{th}$ and $(i+1)^{st}$ amino acids of the protein structure that we consider.    

\vskip 0.2cm
\noindent
$\bullet ~$   If we find only one matching pair of amino acids in the subset $[\Delta\kappa,\Delta\tau]$,
we use the coordinates of its O and C$\beta$ atoms as the predicted coordinates  of the O, C$\beta$ atoms of the $i^{th}$
C$\alpha$ atom.  

\vskip 0.2cm
\noindent
$\bullet ~$  If there are two or more matching pairs,
we use the average value of their O and C$\beta$ coordinates to determine  those of the O and C$\beta$ around the C$\alpha$ of interest.

\vskip 0.2cm
\noindent
$\bullet ~$   If there are no pairs of identical amino acids in the subset $[\Delta\kappa,\Delta\tau]$, we use the average value of {\it all} PDB structures in
this subset to determine the O and C$\beta$ coordinates. 

\vskip 0.2cm
\noindent
$\bullet ~$   Finally, if the subset $[\Delta\kappa,\Delta\tau]$ is empty we search for one from a neighboring subset, first from preceding  $\Delta\tau$, then from following
$\Delta\tau$, then from neighboring $\Delta\kappa$; but such cases are highly exceptional.
 
\vskip 0.2cm
\noindent
{\bf Step 3:~}  We repeat the process for all C$\alpha$ atoms along the chain.  At  the end of the chain there is no $\kappa_{i+1}$, 
thus at the end of the chain  we use only the $\kappa_i$ value in our search.

\vskip 0.2cm 
Our reconstruction algorithm is extremely simple and proceeds very fast computationally, much faster than any of the other three reconstruction programs 
we consider, even though we have not optimized the search algorithm but use a straightforward MatLab code. 

The sector size $\Delta \tau$ can be changed and optimised; here we have chosen 60 sectors, only to exemplify the method.  

The reason why we divide  the original annulus into two using $\kappa=1.2$  in our search algorithm is, that while a torsion angle is determined by four C$\alpha$ atoms,
in the case of a bond angle only three C$\alpha$ are needed; see Figure \ref{fig-2}.  Thus, by engaging the two neighboring bond angle values, 
we employ the full information in all four C$\alpha$ atoms in our search algorithm. 

In Step 2, we calculate the average values of the angles as follows: For the average latitude $\theta_{ave}$ (similarly for $\vartheta_{ave}$) we simply use 
\[
\theta_{ave} \ = \ \frac{1}{N} \sum\limits_{i=k}^N \theta_k
\]
where the summation is over all elements in the given subset  $[\Delta\kappa,\Delta\tau]$. For the average longitude $\phi_{ave}$ (similarly for $\varphi_{ave}$) 
we proceed as follows: We first define
\[
X \ = \ \frac{1}{N} \sum\limits_{i=k}^N \cos \phi_k \ \ \ \ \ \& \ \ \ \ \ Y \ = \ \frac{1}{N} \sum\limits_{i=k}^N \sin \phi_k
\]
and 
\[
R = \sqrt{X^2 + Y^2}
\]
The average value is then obtained from
\[
\cos \phi_{ave} \ = \ \frac{X}{R} \ \ \ \ \ \& \ \ \ \ \ \sin \phi_{ave} \ = \ \frac{Y}{R}
\]

\vskip 0.2cm

\subsection{Algorithm comparisons}

\subsubsection{Direction comparison}

In order to  compare the different reconstruction methods we introduce two unit length 
vectors $\overrightarrow{C\alpha O}$ and $\overrightarrow{C\alpha \beta}$; the former 
points  from the C$\alpha$ atom to the following  peptide plane O atom ($X=$\, O in the sequel), and the latter points from the 
C$\alpha$ atom to its side chain C$\beta$ atom ($X=\beta$ in the sequel). We evaluate these vectors for all 
residues $i=1,...,N$ and for every structure $k=1,...,K$ in the {\it Anton} data. We also evaluate these 
vectors in all of the four reconstruction methods and in the sequel $\rm y=\rm P, R, S, M$ stands 
for {\it Pulchra} (P), {\it Remo} (R), {\it Scwrl4} (S) and the {\it Statistical Method} (M) respectively.

We define $\Theta^{\rm y}_{\rm X}[i,k]$ to be the angle between a vector $\overrightarrow{C\alpha X}$ evaluated
from the {\it Anton} data, and the corresponding vector obtained from the corresponding reconstruction method.
The statistical distribution function for all the values $\Theta^{\rm y}_{\rm X}[i,k]$ measures the overall accuracy
of a given method, for reconstructing the individual O and C$\beta$ atom positions. 

For each of the $k=1,...,K$ {\it Anton} chain structure,  we evaluate the root-mean-square (RMS) value of the angles 
$\Theta^{\rm y}_{\rm X}[i,k]$  by summing over the $N$ individual residues of the chain,
\begin{equation}
{\rm RMS} \,[\Theta^{\rm y}_{\rm X}(k) ] \ = \ \sqrt{ \frac{1}{N} \sum\limits_{i=1}^{N} \left( \Theta^{\rm y}_{\rm X} [i,k]\right)^2 }
\label{rmstheta}
\end{equation}
The distribution density  (\ref{rmstheta}) then measures the overall accuracy of a method, in the reconstruction of the C$\alpha$-O-C$\beta$ chains.  

\subsubsection{Distance comparison}

We also compare the algorithms by evaluating the RMS distance between different atomic positions in
the {\it Anton} trajectory and in the reconstructed trajectory. The RMS distance between  two chain structures 
is evaluated from
\begin{equation}
{\rm RMSD}(X;k) = \sqrt{ \frac{1}{N} \sum\limits_{i=1}^{N} |\mathbf r_{i,X}^{A,k} - \mathbf r_{i,X}^{y,k} |^2 }
\label{RMSD}
\end{equation}
Here $\mathbf r_{i,X}^{A,k} $ is the {\it Anton} data distance between the C$\alpha$
atom and the $X={\rm O}, \, \beta$ atom at residue $i$ in structure $k$, and $\mathbf r_{i,X}^{y,k}$ is the corresponding 
quantity in the $y=\rm P,R,S,M$ reconstructed structure. The probability distribution  of (\ref{RMSD}) is  a complement of 
(\ref{rmstheta}), as a  measure of the overall accuracy of a method in 
the reconstruction of the entire chain in a statistical sense. 

\begin{figure}[h!]         
\centering            
  \resizebox{6.5cm}{!}{\includegraphics[]{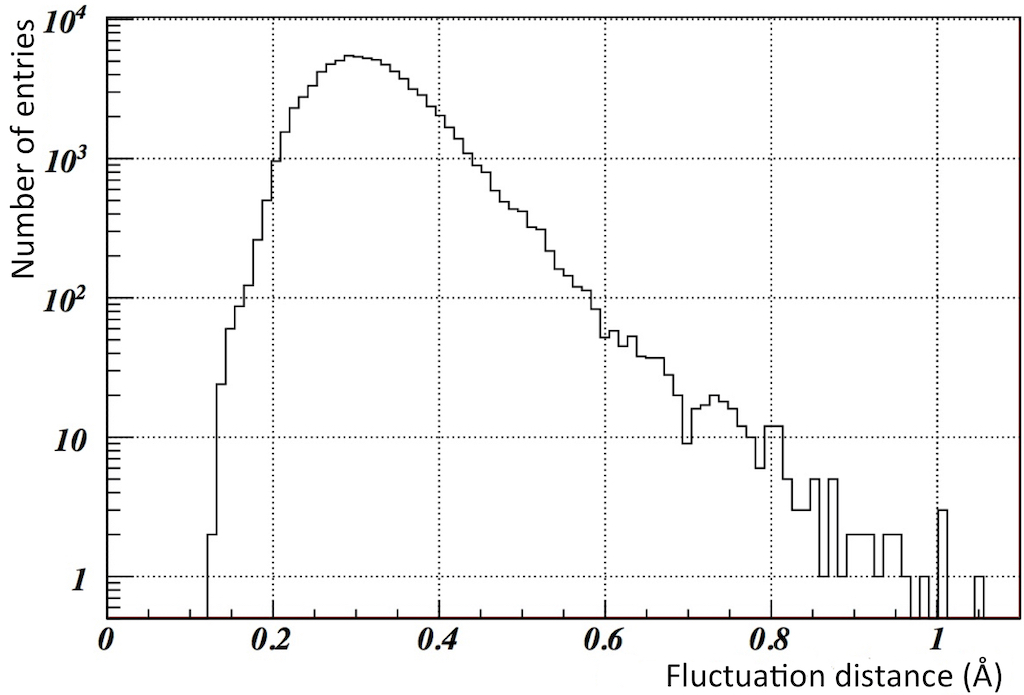}}
\caption {\small  {\it Color online:} 
Combined B-factor fluctuation distances for the O and C$\beta$ atoms
}   
\label{fig-8}    
\end{figure}

For reference, in Figure \ref{fig-8} we present the combined distribution 
of the Debye-Waller fluctuation distances
\[
\sqrt{ < |\mathbf x|^2 > } \ = \ \sqrt{ \frac{B}{8\pi^2}}
\]
for the O and C$\beta$ atoms, that we have evaluated using the $B$-factors in our  1.0 \AA~ PDB pool; note  the logarithmic scale. 
The fluctuation distances are strongly peaked at around 0.3 \AA, and there are practically no structures with a fluctuation distance 
less than 0.12 \AA.
%
%
%%%%%%%%%%%%
%%%%%%%%%%%%%%%
%%%%%%%%%%%%%%%%%%%%%%%%%%%%%%%%%%%%%%%%%%%%%
%%
%%
%%
%%
%%
%%                           FIGURE  8
%%
%%
%%
%%
%%
%% 
%%%%%%%%%%%%%%
%%%%%%%%%%%%%%
%
%
%
%%
%{
%\footnotesize
%
%
%
%
%%%%%%%%%%
%%%%%%%%%%%%%%
%%%%%%%%%%%%%%%

\begin{figure}[h!]         
\centering            
  \resizebox{6.5cm}{!}{\includegraphics[]{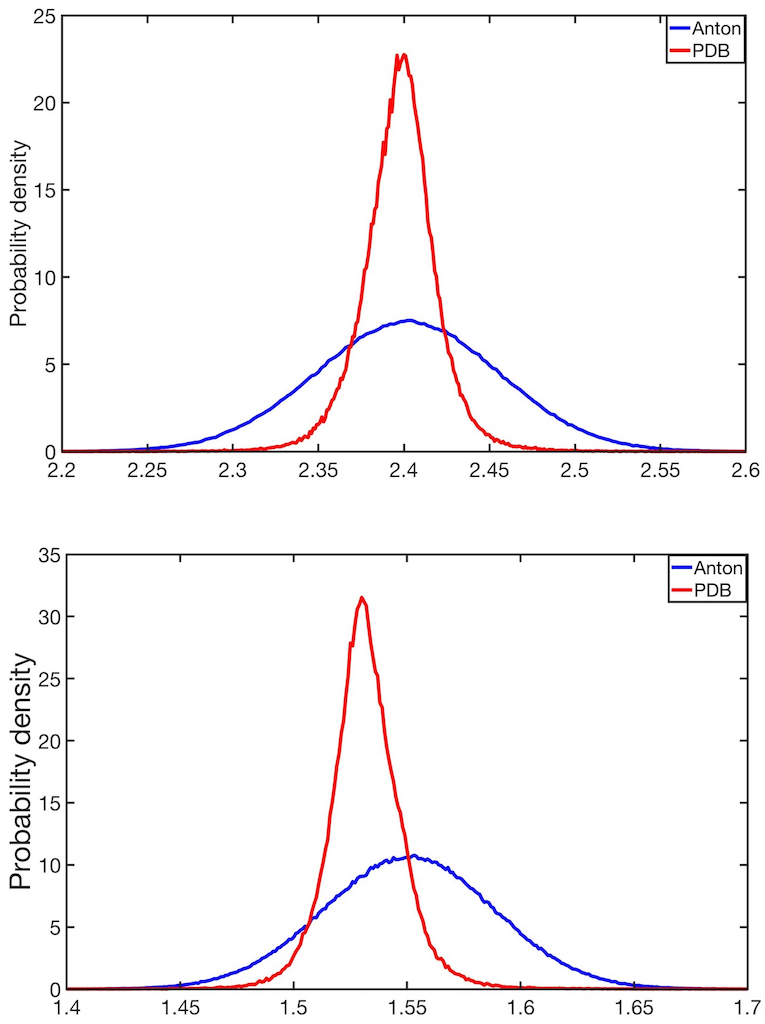}}
\caption {\small  {\it Color online:} 
distribution of distances in \AA ngstr\"om, calculated directly from the coordinates in our PDB data and {\it Anton} data for villin. Top: C$\alpha$-O
Bottom: C$\alpha$-C$\beta$.
}   
\label{fig-9}    
\end{figure}

Similarly, Figure \ref{fig-9} shows the statistical distributions in the  individual C$\alpha$-O and C$\alpha$-C$\beta$ distances that we calculate directly from
the coordinates in our PDB data pool and {\it Anton} data, respectively; only results  for villin are shown as the results for ww-domain are very similar.
%
%
%%%%%%%%%%%%
%%%%%%%%%%%%%%%
%%%%%%%%%%%%%%%%%%%%%%%%%%%%%%%%%%%%%%%%%%%%%
%%
%%
%%
%%
%%
%%                           FIGURE  9
%%
%%
%%
%%
%%
%% 
%%%%%%%%%%%%%%
%%%%%%%%%%%%%%
%
%
%
%%
%{
%\footnotesize
%
%
%
%
%%%%%%%%%%
%%%%%%%%%%%%%%
%%%%%%%%%%%%%%%
In the following we shall not refine the C$\alpha$-O and the C$\alpha$-C$\beta$ distances, in our statistical method, the differences are in any case
minor. Instead we simply use the average PDB distance values 
\[
\begin{matrix}
\Delta r \ = \ 2.40 \ {\rm \AA} \ \ \ \ \ \ \ {\rm For ~ C\alpha-O ~ distance} \\
\hskip 0.15cm \Delta r \ = \ 1.53 \ {\rm \AA} \ \ \ \ \ \ \ {\rm For ~ C\alpha-C\beta ~ distance} \\
\end{matrix}
\]
in our {\it Statistical Method} reconstruction.  It is natural to allocate the  larger variance in the case of {\it Anton} data in Figures \ref{fig-9} 
to temperature fluctuations.

\section{Results}

\subsection{Peptide plane O atoms}

We start with the peptide plane O atom distributions in the {\it Anton} data.   We compare the {\it Anton} results 
shown in Figure \ref{fig-4}, with results from {\it Pulchra} , {\it Remo} and  {\it Statistical Method}. 

\subsubsection{Frenet Spheres}

\begin{figure}[h!]         
\centering            
  \resizebox{6.cm}{!}{\includegraphics[]{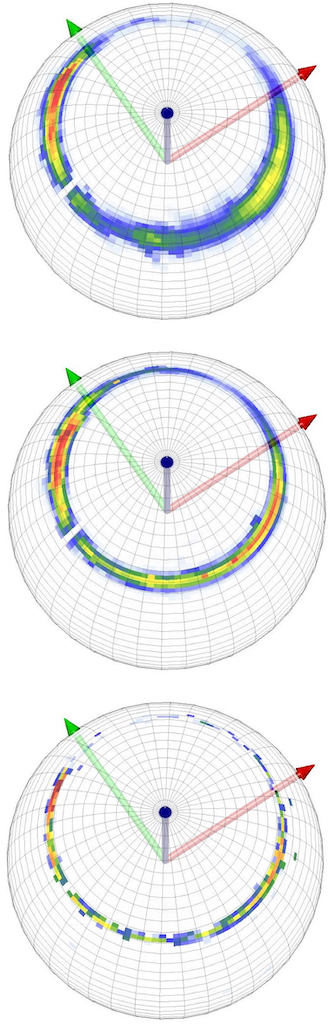}}
\caption {\small  {\it Color online:} 
Top: Reconstructed peptide plane O distribution for villin  
on Frenet sphere $\mathbb S^2$.  Top: {\it Pulchra} Middle: {\it Remo} Bottom: {\it Statistical Method}.
}   
\label{fig-10}    
\end{figure}

In Figures \ref{fig-10} we present  the reconstructed O atom distributions on a Frenet sphere $\mathbb S^2$,  in the case of villin.
The Figures display all the reconstructed O atom coordinates ($\theta,\phi$) for all the C$\alpha$ atoms of all   {\it Anton}
trajectories that we obtain using {\it Pulchra}, {\it Remo} and {\it Statistical Method} respectively.

%
%
%%%%%%%%%%%%
%%%%%%%%%%%%%%%
%%%%%%%%%%%%%%%%%%%%%%%%%%%%%%%%%%%%%%%%%%%%%
%%
%%
%%
%%
%%
%%                           FIGURE  10
%%
%%
%%
%%
%%
%% 
%%%%%%%%%%%%%%
%%%%%%%%%%%%%%
%
%
%
%%
%{
%\footnotesize
% 
%
%
%
%%%%%%%%%%
%%%%%%%%%%%%%%
%%%%%%%%%%%%%%%
Overall,  all three distributions reproduce  well the O atom  villin distribution of the {\it Anton} trajectory in  Figure \ref{fig-4} (top). In particular, the regular 
$\alpha$-helical and  $\beta$-stranded regions are clearly identifiable. We observe that both {\it Pulchra} and {\it Remo} distributions are slightly wider than the
PDB distribution, we propose that this reflects mainly the presence of  thermal effects in the {\it Anton} data, as captured by these two methods.  
On the other hand, the {\it Statistical Method} distribution is more concentrated. This  is expected since the data pool is a subset of the PDB data shown in
Figure \ref{fig-4} (top), which have all been measured at very low temperature values.

We remind that the color coding is always relative but equal, in all the cases that we display, and intensity increases from no-entry white to low density 
blue, and towards high density red.

\begin{figure}[h!]         
\centering            
  \resizebox{6cm}{!}{\includegraphics[]{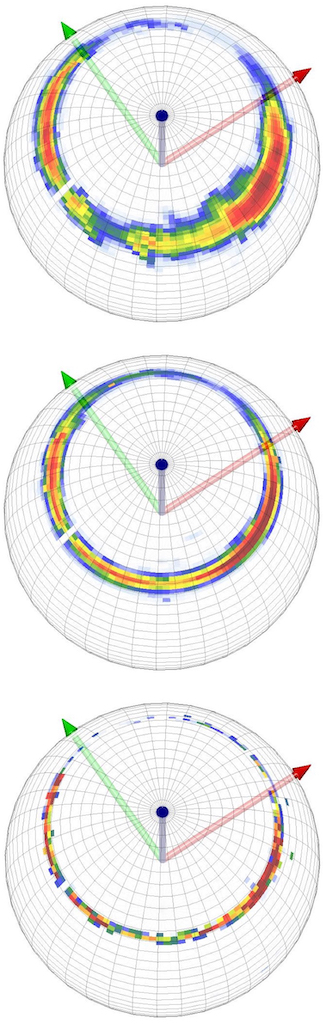}}
\caption {\small  {\it Color online:} 
Top: Reconstruction of peptide plane O distribution  
on the sphere $\mathbb S_\alpha^2$ for ww-domain.  Top: {\it Pulchra} Middle: {\it Remo} Bottom: {\it Statistical Method}.}
\label{fig-11}    
\end{figure}

In Figures \ref{fig-11} we present  the reconstructed O distributions  in the case of ww-domain. 
Again, all three distributions are very much in line with the {\it Anton} data of
Figure \ref{fig-4} (bottom). We observe some  fragmentation and excess (thermal) data spreading in the case of {\it Pulchra}. The {\it Remo} distribution 
also displays thermal spreading while the {\it Statistical Method} distribution is again more concentrated, as expected since it forms a
subset of the low temperature PDB data. 

We now  analyze the reconstruction results for the peptide plane O atoms  in more detail, using the methods of Section II F.
%
%
%%%%%%%%%%%%
%%%%%%%%%%%%%%%
%%%%%%%%%%%%%%%%%%%%%%%%%%%%%%%%%%%%%%%%%%%%%
%%
%%
%%
%%
%%
%%                           FIGURE  11
%%
%%
%%
%%
%%
%% 
%%%%%%%%%%%%%%
%%%%%%%%%%%%%%
%
%
%
%%
%{
%\footnotesize
%
%
%
%
%%%%%%%%%%
%%%%%%%%%%%%%%
%%%%%%%%%%%%%%%

\subsubsection{Individual  angular probability densities for peptide planes}

\begin{figure}[h!]         
\centering            
  \resizebox{6cm}{!}{\includegraphics[]{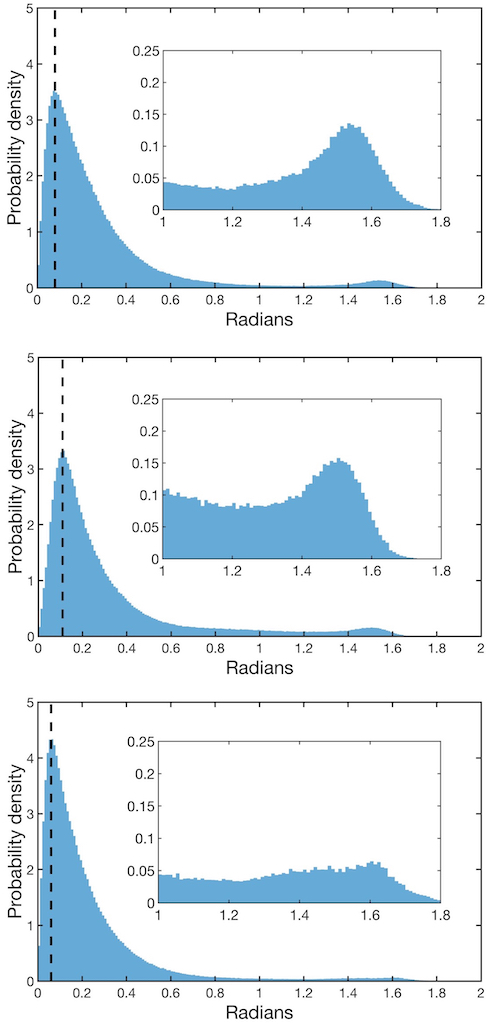}}
\caption {\small  {\it Color online:} 
Probability density distribution for all individual angles $\Theta^{\rm y}_{\rm O}[i,k]$ in the case of {\it Anton} villin trajectories.
Top: {\it Pulchra} Middle: {\it Remo} Bottom: {\it Statistical Method}.
}
\label{fig-12}    
\end{figure}

\begin{figure}[h!]         
\centering            
  \resizebox{6cm}{!}{\includegraphics[]{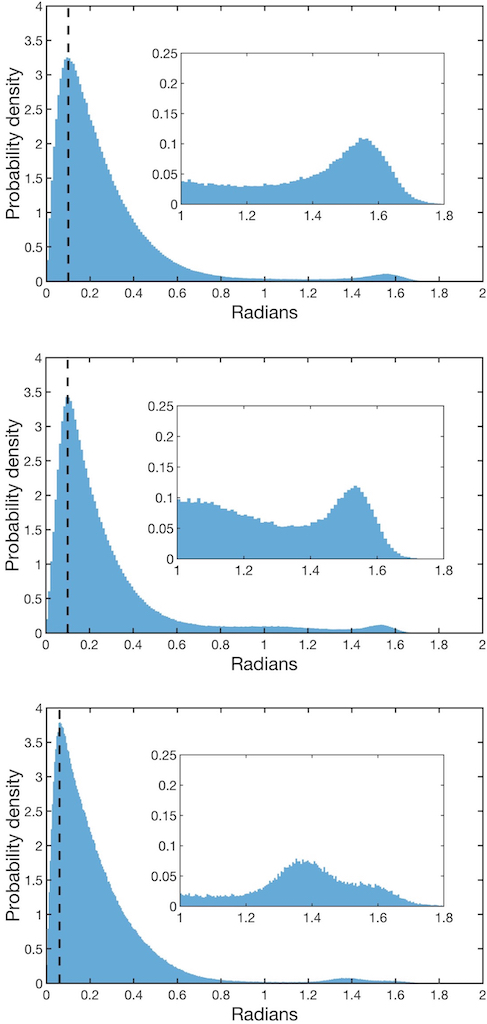}}
\caption {\small  {\it Color online:} 
Probability density distribution  for all angles $\Theta^{\rm y}_{\rm O}[i,k]$ along the  {\it Anton} ww-domain trajectories.
Top: {\it Pulchra} Middle: {\it Remo} Bottom: {\it Statistical Method}.
}
\label{fig-13}    
\end{figure}

In Figures \ref{fig-12} and \ref{fig-13} we show the normalized probability density distributions
for all the individual angles $\Theta^{\rm y}_{\rm O}[i,k]$ along the {\it Anton} trajectories for {\it Pulchra}, {\it Remo} and the {\it Statistical method} 
in the case of villin and ww-domain O atoms, respectively.

%
%
%%%%%%%%%%%%
%%%%%%%%%%%%%%%
%%%%%%%%%%%%%%%%%%%%%%%%%%%%%%%%%%%%%%%%%%%%%
%%
%%
%%
%%
%%
%%                           FIGURE  12
%%
%%
%%
%%
%%
%% 
%%%%%%%%%%%%%%
%%%%%%%%%%%%%%
%
%
%
%%
%{
%\footnotesize
%
%
%
%
%%%%%%%%%%
%%%%%%%%%%%%%%
%%%%%%%%%%%%%%%

%
%
%%%%%%%%%%%%
%%%%%%%%%%%%%%%
%%%%%%%%%%%%%%%%%%%%%%%%%%%%%%%%%%%%%%%%%%%%%
%%
%%
%%
%%
%%
%%                           FIGURE  13
%%
%%
%%
%%
%%
%% 
%%%%%%%%%%%%%%
%%%%%%%%%%%%%%
%
%
%
%%
%{
%\footnotesize
%
%
%
%
%%%%%%%%%%
%%%%%%%%%%%%%%
%%%%%%%%%%%%%%%

In both {\it Pulchra} and {\it Remo}  the individual $\Theta^{\rm y}_{\rm O}[i,k]$ values peak near the small value $\Theta_{max} \approx  0.1$ (rad). For 
the {\it Statistical Method}  the peak is located at the even smaller  value $\Theta_{max}  \approx 0.06$ (rad), both in the case of villin and ww-domain. 

From Figure \ref{fig-9} we learn that the average PDB distance between C$\alpha$ and O is around 2.4  \AA. An angular deviation of $\Theta_{max}  \approx 0.06$ then 
corresponds to a distance deviation $\sim 0.14$ \AA ~ which  is smaller than the B-factor fluctuation distances in Figure \ref{fig-8}  and more in line with  
the (apparently purely thermal) distance deviations we observe  in Figures \ref{fig-9}.

We conclude that  each of the three methods can reconstruct the individual angular positions
of the dynamical {\it Anton} O atoms with very high precision. Moreover, despite its simplicity 
the {\it Statistical Method}  performs even better than both {\it Pulchra} and {\it Remo}. 

In each of the probability distributions of Figures \ref{fig-12}, \ref{fig-13} we observe enhanced accumulation of data near $\Theta \approx \pi/2$; 
the inserts show the probability densities for $\Theta$-values above 1.0 (rad). We recall
 our  interpretation of the Frenet sphere
O distribution as the base of a cone, with the apex at the origin; 
the vertex angle has a value very close to $\pi/2$. Thus the $\Theta \approx \pi/2$  secondary peak
corresponds to a $\phi \sim$180 degree rotation around the (blue) $\mathbf t$-vector in the  Figures \ref{fig-10}, \ref{fig-11}.
We observe  that a rotation of the longitude $\phi$ by  $\sim 180^{\rm o}$ exchanges  the $\alpha$-helical and $\beta$-stranded regions according to top Figure \ref{fig-3}. 

\subsubsection{Angular probability densities for entire  chains}

\begin{figure}[h!]         
\centering            
  \resizebox{6.0cm}{!}{\includegraphics[]{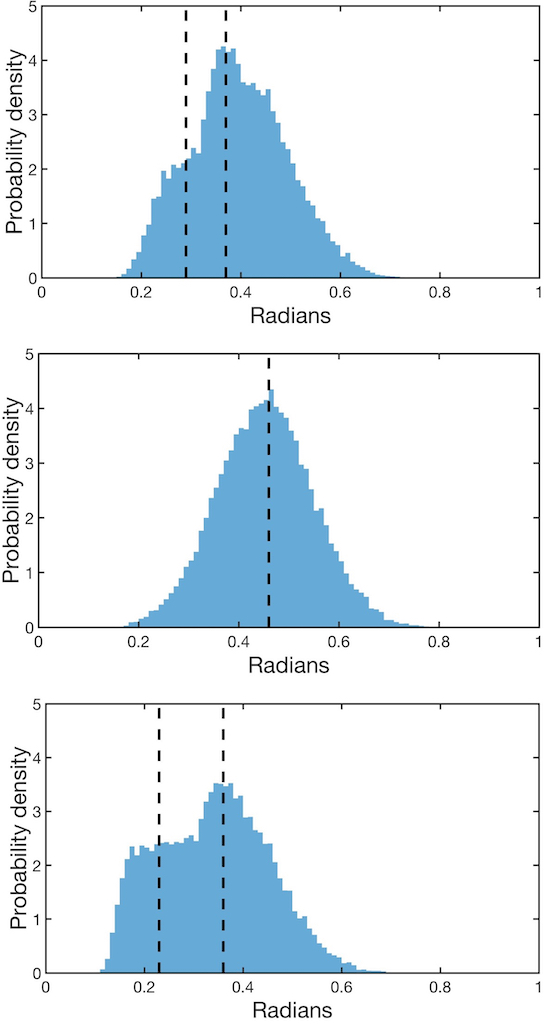}}
\caption {\small  {\it Color online:} 
The probability density distributions (\ref{rmstheta}) for villin. 
Top: {\it Pulchra} Middle: {\it Remo} Bottom: {\it Statistical Method}.
}   
\label{fig-14}    
\end{figure}

\begin{figure}[h!]         
\centering            
  \resizebox{6.5cm}{!}{\includegraphics[]{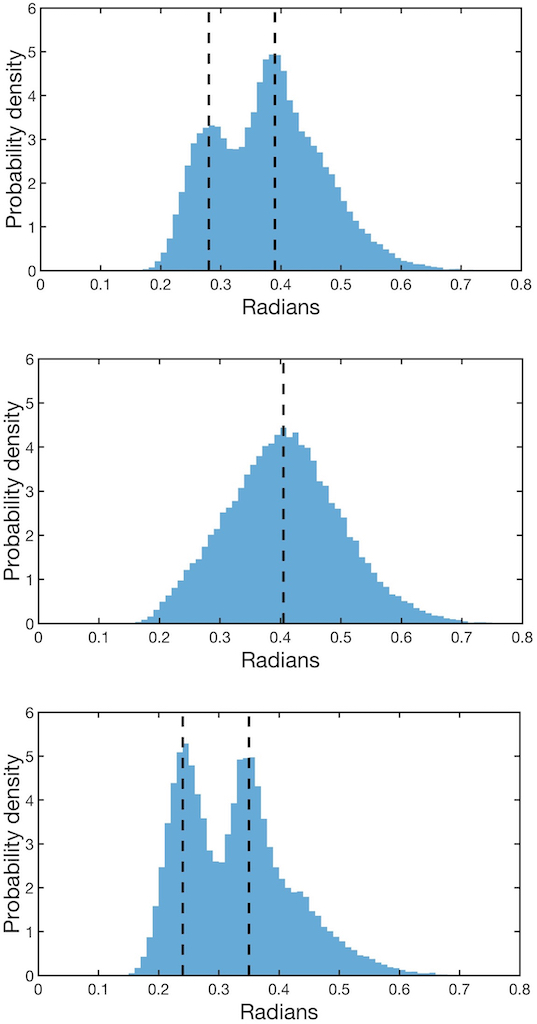}}
\caption {\small  {\it Color online:} 
The probability density distributions (\ref{rmstheta}) for ww-domain. 
Top: {\it Pulchra} Middle: {\it Remo} Bottom: {\it Statistical Method}.
}
\label{fig-15}    
\end{figure}

In Figures \ref{fig-14}, \ref{fig-15} we show the probability density distributions (\ref{rmstheta}) for
{\it Pulchra}, {\it Remo} and {\it Statistical Method}, in the case of villin and ww-domain O atoms, respectively. We remind  that the value of  (\ref{rmstheta}) is a 
measure for  the accuracy of the reconstruction, in the case of the entire chain.

%
%
%%%%%%%%%%%%
%%%%%%%%%%%%%%%
%%%%%%%%%%%%%%%%%%%%%%%%%%%%%%%%%%%%%%%%%%%%%
%%
%%
%%
%%
%%
%%                           FIGURE  14
%%
%%
%%
%%
%%
%% 
%%%%%%%%%%%%%%
%%%%%%%%%%%%%%
%
%
%
%%
%{
%\footnotesize
%
%
%
%
%%%%%%%%%%
%%%%%%%%%%%%%%
%%%%%%%%%%%%%%%

%
%
%%%%%%%%%%%%
%%%%%%%%%%%%%%%
%%%%%%%%%%%%%%%%%%%%%%%%%%%%%%%%%%%%%%%%%%%%%
%%
%%
%%
%%
%%
%%                           FIGURE  15
%%
%%
%%
%%
%%
%% 
%%%%%%%%%%%%%%
%%%%%%%%%%%%%%
%
%
%
%%
%{
%\footnotesize
%
%
%
%
%%%%%%%%%%
%%%%%%%%%%%%%%
%%%%%%%%%%%%%%%
%
%

In each case, the reconstructed chains appear to be very close to the original {\it Anton}  chains, both in the case of  villin and ww-domain. 
%The deviations are slightly above the average B-factor fluctuation distance  of Figure \ref{fig-8}.
The {\it Statistical Method} performs best and the results for {\it Pulchra} are quite similar,  but for {\it Remo} we observe a clear deviation from
the {\it Statistical method}; the results from the latter are visibly better.  

Note that  in the case of  both {\it Pulchra} and {\it Statistical Method},  both the villin and the ww-domain profiles appear to 
resemble a combination of two distinct Gaussian distributions. On the other hand,
in the case of {\it Remo} the distribution is  like a single Gaussian (thermal spread),  in both cases.

\subsubsection{RMSD probability densities for entire chains}

\begin{figure}[h!]         
\centering            
  \resizebox{6.5cm}{!}{\includegraphics[]{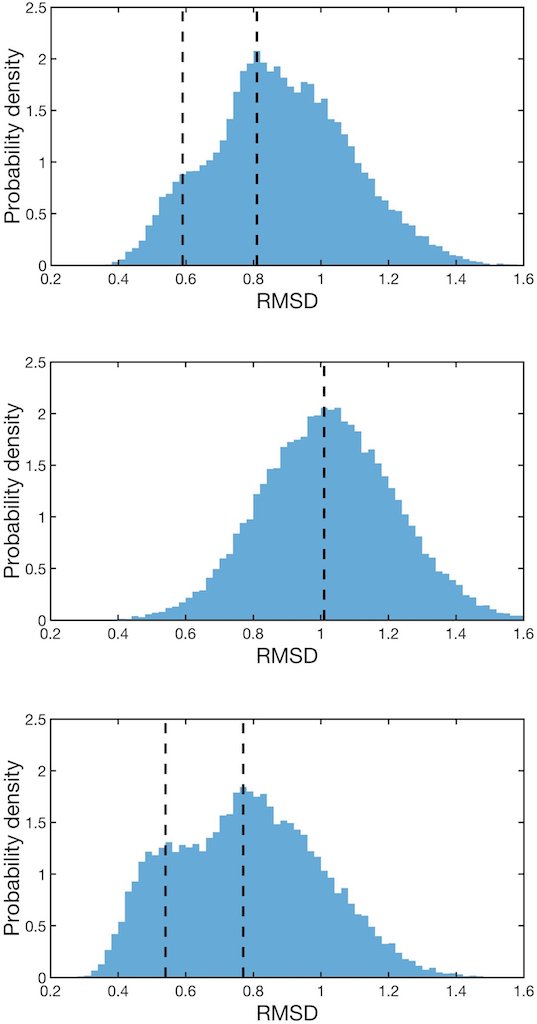}}
\caption {\small  {\it Color online:} 
The probability density distributions for the RMS distance (\ref{RMSD}) over the O atoms in the case of villin. 
Top: {\it Pulchra} Middle: {\it Remo} Bottom: {\it Statistical Method}.
}
\label{fig-16}    
\end{figure}

\begin{figure}[h!]         
\centering            
  \resizebox{6.5cm}{!}{\includegraphics[]{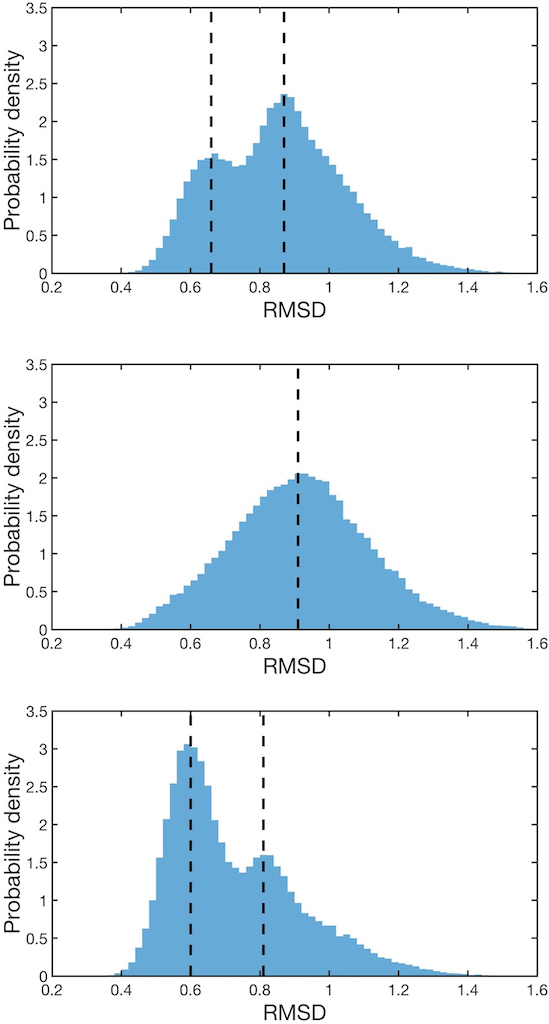}}
\caption {\small  {\it Color online:} 
The probability density distributions for the RMS distance (\ref{RMSD}) over the O atoms in the case of ww-domain. 
Top: {\it Pulchra} Middle: {\it Remo} Bottom: {\it Statistical Method}.
}
\label{fig-17}    
\end{figure}

In Figures \ref{fig-16}, \ref{fig-17} we show the probability density distributions for the RMS distances  (\ref{RMSD}), 
evaluated for the entire C$\alpha$-O reconstructed chains that we obtain 
using {\it Pulchra}, {\it Remo} and {\it Statistical Method} in the case of  villin and ww-domain {\it Anton} trajectories.

%%%%%%%%%%%%
%%%%%%%%%%%%%%%
%%%%%%%%%%%%%%%%%%%%%%%%%%%%%%%%%%%%%%%%%%%%%
%%
%%
%%
%%
%%
%%                           FIGURE  16
%%
%%
%%
%%
%%
%% 
%%%%%%%%%%%%%%
%%%%%%%%%%%%%%
%
%
%
%%
%{
%\footnotesize
%
%
%
%
%%%%%%%%%%
%%%%%%%%%%%%%%
%%%%%%%%%%%%%%%

%
%
%%%%%%%%%%%%
%%%%%%%%%%%%%%%
%%%%%%%%%%%%%%%%%%%%%%%%%%%%%%%%%%%%%%%%%%%%%
%%
%%
%%
%%
%%
%%                           FIGURE  17
%%
%%
%%
%%
%%
%% 
%%%%%%%%%%%%%%
%%%%%%%%%%%%%%
%
%
%
%%
%{
%\footnotesize
%
%
%
%
%%%%%%%%%%
%%%%%%%%%%%%%%
%%%%%%%%%%%%%%%

Again, the reconstruction by the {\it Statistical Method} is closest to the original  {\it Anton} result.
The difference to {\it Pulchra} is very small  but there is a more visible difference to {\it Remo}.

Even though all the RMS distances between the O-chains shown in Figures \ref{fig-16} and \ref{fig-17} are small, they are slightly larger than what we can 
expect on the basis of the individual (O and C$\beta$) atom B-factor fluctuation distances 
in Figure \ref{fig-8}. In line with Figures \ref{fig-14}, \ref{fig-15}, both {\it Pulchra} and {\it Statistical Method} distributions again exhibit a  double Gaussian profile;
In the case of {\it Remo} the distributions form a single Gaussian which is more  in line with a thermal spread, except that the peak is close to 1.0 \AA ~ which is a somewhat
large value in comparison to the {\it Statistical Method} where  the two peaks are slightly below values 
0.6 \AA ~ and 0.8 \AA ~ {\it i.e.} close to the covalent radius  $\sim 0.66$ \AA ~ of O atom.   But  even in the case of {\it Remo},  the deviation distances can be  
considered  to be quite  small and we consider the results to be good.

\subsubsection{Peptide plane flip}

In Figures \ref{fig-12} and \ref{fig-13} we have noted an accumulation of entries near $\Theta \sim \pi/2$,  corresponding to a  $\sim$180 degree rotation of the entire
probability distribution around the $\mathbf t$ vector. Consequently these entries contribute 
maximally to the deviations between the {\it Anton} chain and the reconstructed
chains, in terms of statistical distributions. We note that $\Theta \sim \pi/2$ corresponds to $\sim 3.4$ \AA ~ in terms of spatial distance.

\begin{figure}[h!]         
\centering            
  \resizebox{6.5cm}{!}{\includegraphics[]{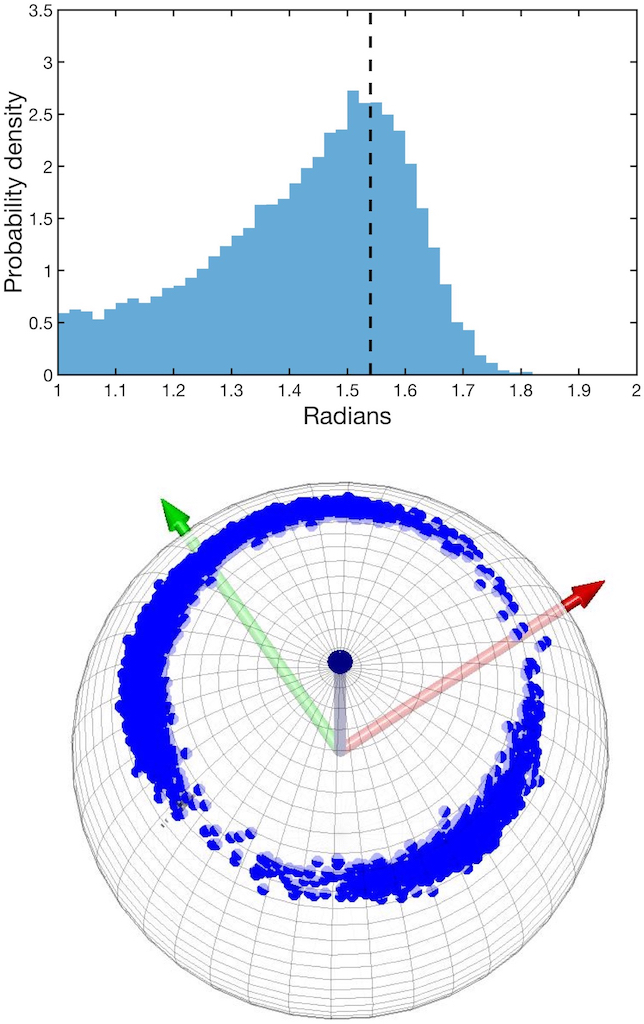}}
\caption {\small  {\it Color online:} 
Top: The distribution of average value of angular deviations for the reconstructed entries with $\Theta > 1.0$ (rad)
in Figures \ref{fig-12} (villin). Bottom: The distribution of these entries on the Frenet sphere (ww-domain).
}
\label{fig-18}    
\end{figure}

We consider only those {\it Anton}  entries for which $\Theta > 1.0$  simultaneously, in  {\it all} of the three reconstruction methods; the spatial distance for two entries
that are $\Theta \sim 1.0$ apart is $\sim$2.3 \AA ~ which is way above the range of B-factor fluctuations in Figure \ref{fig-8}. Thus, the $\Theta > 1.0$ entries  that are common
to all three methods should  be very little  prone to method dependent fallacies, these entries should correspond to definite deviations in {\it Anton} data from PDB structures. 
There are a total of around 6.000 such 
entries, this corresponds to a mere 0.6 per cent of the total entries that we have analyzed. We evaluate the average values of $\Theta$ for all these  entries. 
The probability distribution for these average values is concentrated  very close to $\Theta = \pi/2$ as shown in Figure \ref{fig-18} (top). The entries
are also distributed quite evenly around the O-circle on the Frenet sphere (Figure \ref{fig-18}) (bottom), they seem to appear quite randomly 
during the {\it Anton} time evolution, and correspond to very sudden and short-lived 180 degree back-and-forth rotations (flips) of the entire peptide 
plane, around the virtual bond that connect two consecutive C$\alpha$ atoms. 

From the available {\it Anton} data, we are unable to conclude
whether these peptide plane flips are  genuine physical effects with an important role in protein folding, or whether they are mere simulation artifacts, or whether they are effects 
that are specific  to the CHARMM22$^\star$ force field \cite{Piana-2011}.  For example,  
it appears that in the {\it Anton} simulations the peptide plane N-H covalent bond lengths  are constrained to have a fixed value. That may affect the stability of peptide planes, causing
them to flip by 180 degrees. 

To properly understand the character of these 
peptide plane  flips one needs to perform more detailed all atom simulations, 
presumably with shorter time steps than used in {\it Anton} simulations and with no length constraints on the N-H covalent bonds.   
%
%
%%%%%%%%%%%%
%%%%%%%%%%%%%%%
%%%%%%%%%%%%%%%%%%%%%%%%%%%%%%%%%%%%%%%%%%%%%
%%
%%
%%
%%
%%
%%                           FIGURE  18
%%
%%
%%
%%
%%
%% 
%%%%%%%%%%%%%%
%%%%%%%%%%%%%%
%
%
%
%%
%{
%\footnotesize
%
%
%
%
%%%%%%%%%%
%%%%%%%%%%%%%%
%%%%%%%%%%%%%%%

\subsubsection{Double Gaussians}

In Figures \ref{fig-14}-\ref{fig-17} we have observed the presence of  a double  Gaussian peak structures, in the 
{\it Pulchra} and {\it Statistical Method} distributions. We now proceed to try and identify the cause. We present  a detailed 
analysis of the double Gaussian structures in the case of the {\it Statistical Method } RMSD distributions in Figures \ref{fig-16} and \ref{fig-17};
the analysis in the other cases is similar, with similar conclusions.

In \cite{Lindorff-2011} the following quantity 
\begin{equation}
Q(t) \ = \ \frac{   
\sum_{i=1}^{N_{res}} 
 \sum_{j=1}^{N_i} 
  \ \left[  
  1 + e^{ 10 ( d_{ij}(t) - (d^0_{ij}+1))}
  \right ]^{-1}
}
{  \sum_{i=1}^{N_{res}} N_i }
\label{Qan}
\end{equation}
has been introduced, to characterize the distance of a dynamical chain at time $t$,  
to an experimentally measured  folded state. Here $N_i$ is the number of contacts of residue $i$ along the chain
as defined in \cite{Lindorff-2011}, $d_{ij}(t)$ is the distance in \AA ~ between the C$\alpha$ atoms of residues $i$ and $j$ at time $t$ and
$d_{ij}^0$ is the same distance in the natively folded, crystallographic structure. According to  \cite{Lindorff-2011} a structure with $Q>0.9$ is  
folded and a structure with $Q<0.1$ is  unfolded. 

\begin{figure}[h!]         
\centering            
  \resizebox{8.5cm}{!}{\includegraphics[]{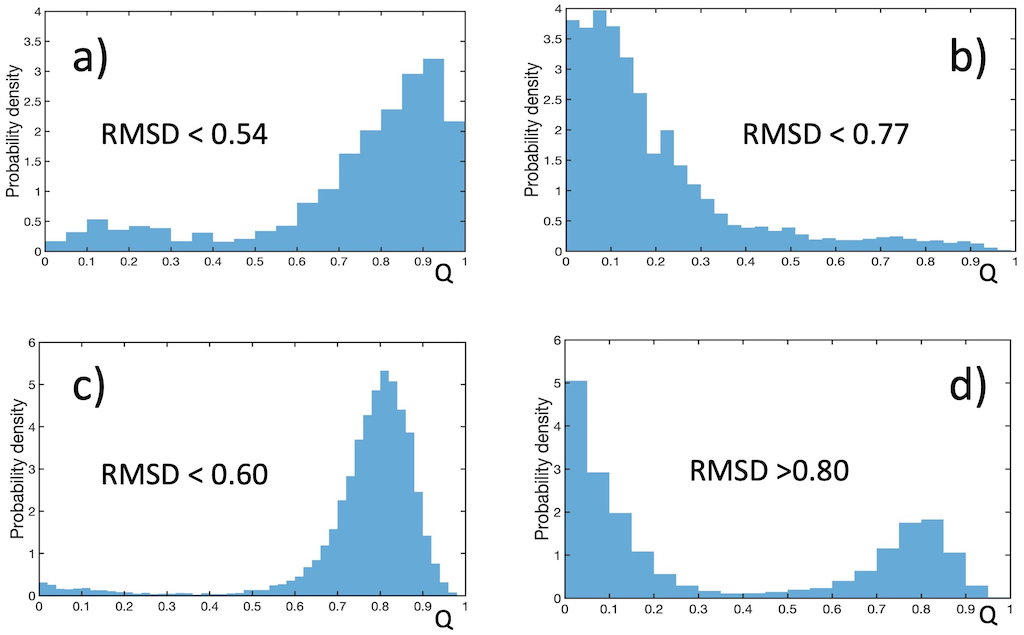}}
\caption {\small  {\it Color online:} 
{\it Statistical Method} probability distributions for the $Q$-values (\ref{Qan}), corresponding to the Gaussian peaks in Figures \ref{fig-16} and \ref{fig-17}
a) low-RMSD villin, b) high-RMSD villin,  c) low-RMSD ww-domain, d) high-RMSD ww-domain. 
}
\label{fig-19}    
\end{figure}

Consider now the two {\it Statistical Method} peaks in the villin distribution Figure \ref{fig-16} and in the ww-domain distribution Figure \ref{fig-17}.
In both cases, we analyze separately the low RMSD subsets with values below 0.54 \AA~ in villin and below 0.6 \AA~ in ww-domain, and the high RMSD
subsets with RMSD values above  0.77 \AA~ in villin and above 0.8 \AA~ in ww-domain. These four subsets are identified by the dashed lines in the Figures.
In Figures \ref{fig-19} we show the probability density distributions for the values of (\ref{Qan})  that we evaluate for these subsets. 
Both in the case of villin and ww-domain the low-RMSD peak corresponds to large values of $Q$ which is characteristic to near-folded state,  
while the large-RMSD peak corresponds to predominantly small values of $Q$ which are characteristic to unfolded states.  
%
%
%%%%%%%%%%%%
%%%%%%%%%%%%%%%
%%%%%%%%%%%%%%%%%%%%%%%%%%%%%%%%%%%%%%%%%%%%%
%%
%%
%%
%%
%%
%%                           FIGURE  19
%%
%%
%%
%%
%%
%% 
%%%%%%%%%%%%%%
%%%%%%%%%%%%%%
%
%
%
%%
%{
%\footnotesize
%
%
%
%
%%%%%%%%%%
%%%%%%%%%%%%%%
%%%%%%%%%%%%%%%

\subsection{Side chain C$\beta$ atoms}

We proceed to describe results from  the C$\beta$ atom reconstruction.  Due to the 
high accuracy in all the methods we use, the presentation is less detailed than in the case of peptide plane O atoms:  The differences to {\it Anton} simulation results are minuscule. 

We investigate reconstruction results in four approaches: {\it Pulchra}, {\it Remo}, {\it Pulchra+Scwrl4}, and  our
{\it Statistical method}. We note that {\it Remo} commonly constructs  the side chain atoms using {\it Scwrl} while {\it Pulchra} employs its own side chain reconstruction.
Thus we  have added a {\it Pulchra+Scwrl4} combination, where the peptide planes are first constructed with {\it Pulchra} and then side chains are constructed with {\it Scwrl4}. 
This combination should be of interest,  since we have found that  {\it Pulchra} performs slightly better  than {\it Remo}  in the reconstruction of the {\it Anton} peptide plane O atom positions. 

\subsubsection{Frenet spheres}

\begin{figure}[h!]         
\centering            
  \resizebox{7.cm}{!}{\includegraphics[]{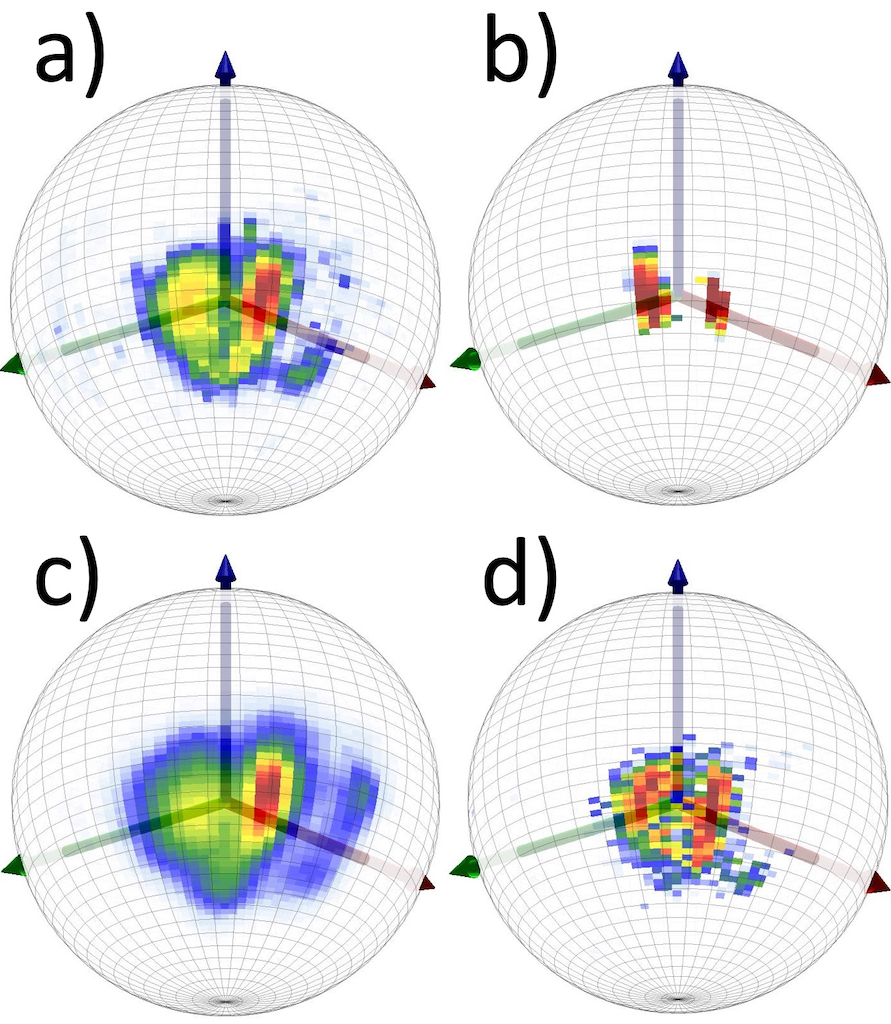}}
\caption {\small  {\it Color online:} 
Reconstructed side chain C$\beta$  distribution for villin on the Frenet sphere $\mathbb S^2_\alpha$.  Figure a) {\it Pulchra}, Figure b) {\it Remo}, Figure 
c) {\it Pulchra+Scwrl4}, Figure d) {\it Statistical Method} 
}
\label{fig-20}    
\end{figure}

\begin{figure}[h!]         
\centering            
  \resizebox{7.5cm}{!}{\includegraphics[]{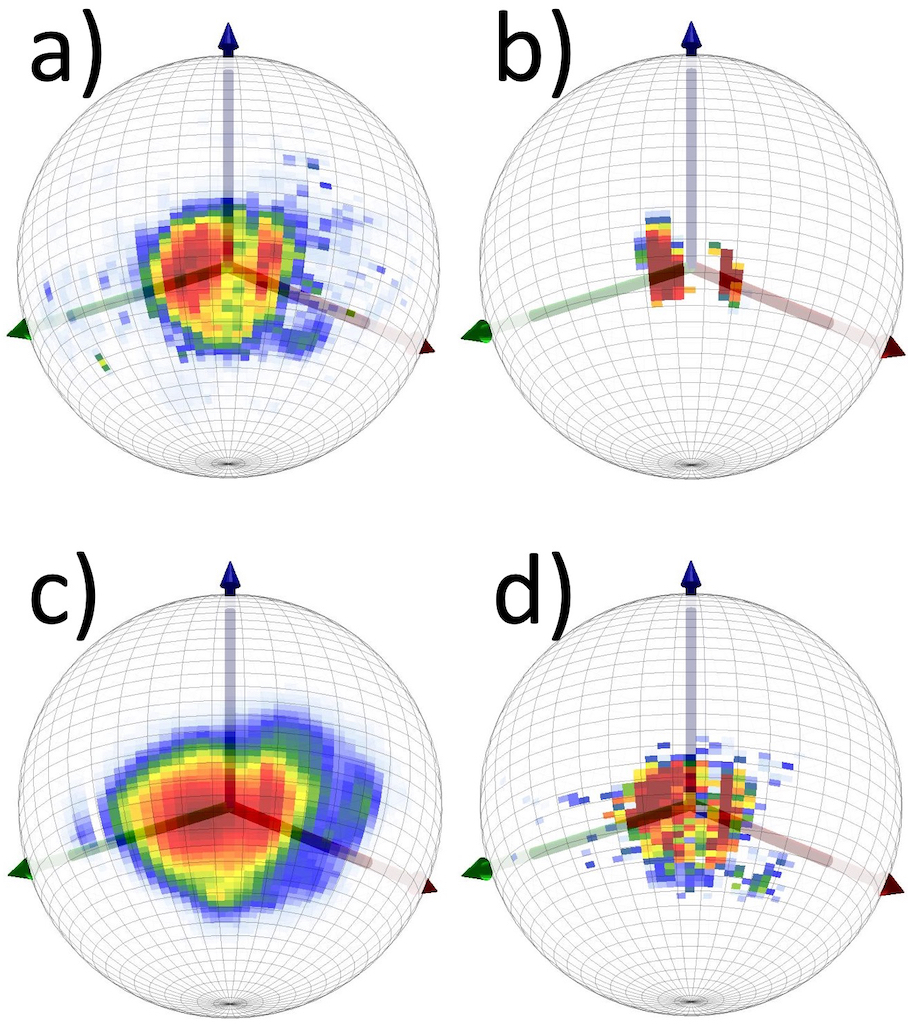}}
\caption {\small  {\it Color online:} 
Reconstructed side chain C$\beta$  distribution for  ww-domain on the Frenet sphere $\mathbb S^2_\alpha$.  Figure a) {\it Pulchra}, Figure b) {\it Remo}, Figure 
c) {\it Pulchra+Scwrl4}, Figure d) {\it Statistical Method} 
}
\label{fig-21}    
\end{figure}

In Figures \ref{fig-20} and \ref{fig-21}  we have the Frenet sphere probability distributions for the four methods, in the case of villin and ww-domain respectively.

%
%
%%%%%%%%%%%%
%%%%%%%%%%%%%%%
%%%%%%%%%%%%%%%%%%%%%%%%%%%%%%%%%%%%%%%%%%%%%
%%
%%
%%
%%
%%
%%                           FIGURE  20
%%
%%
%%
%%
%%
%% 
%%%%%%%%%%%%%%
%%%%%%%%%%%%%%
%
%
%
%%
%{
%\footnotesize
%
%
%
%
%%%%%%%%%%
%%%%%%%%%%%%%%
%%%%%%%%%%%%%%%

%
%
%%%%%%%%%%%%
%%%%%%%%%%%%%%%
%%%%%%%%%%%%%%%%%%%%%%%%%%%%%%%%%%%%%%%%%%%%%
%%
%%
%%
%%
%%
%%                           FIGURE  21
%%
%%
%%
%%
%%
%% 
%%%%%%%%%%%%%%
%%%%%%%%%%%%%%
%
%
%
%%
%{
%\footnotesize
%
%
%
%
%%%%%%%%%%
%%%%%%%%%%%%%%
%%%%%%%%%%%%%%%

{\it Pulchra} reconstructs quite accurately the {\it Anton}  distributions of the  C$\beta$ atoms, shown in Figure \ref{fig-5}. In particular, it 
reconstructs the region of left-handed $\alpha$-helices. {\it Pulchra} also broadens the overall shape of the  distribution in a manner which, at least
superficially,  appears to account for the thermal fluctuations that are present in the {\it Anton} distributions.

{\it Pulchra} and {\it Scwrlt4} combination increases the spread of the C$\beta$ distributions, there is now an even
better resemblance between the {\it Anton} distributions with 
the (perceived) thermal fluctuations. 

{\it Remo} reconstruct the C$\beta$ distributions in terms of very concentrated distributions that are highly peaked at the C$\alpha$ and C$\beta$ regions, with
no observable thermal spreading: 
{\it Remo} appears to reconstruct the C$\beta$ positions in a very straightforward two-stage fashion. In particular,  there  is a marked difference between the
{\it Pulchra+Scwrl4} and {\it Remo} distributions even though  {\it Remo} apparently uses {\it Scwrl} for side chain reconstruction  \cite{Li-2009}.

{\it Statistical Method} selects a subset of the full PDB distribution in  Figure \ref{fig-3} (bottom), as expected. Since the PDB data has been mostly measured
at liquid nitrogen temperatures below  $\sim$77 K, the distributions do not display  thermal spreading.

\subsubsection{Individual angular probability densities for side chains}

\begin{figure}[h!]         
\centering            
  \resizebox{6.5cm}{!}{\includegraphics[]{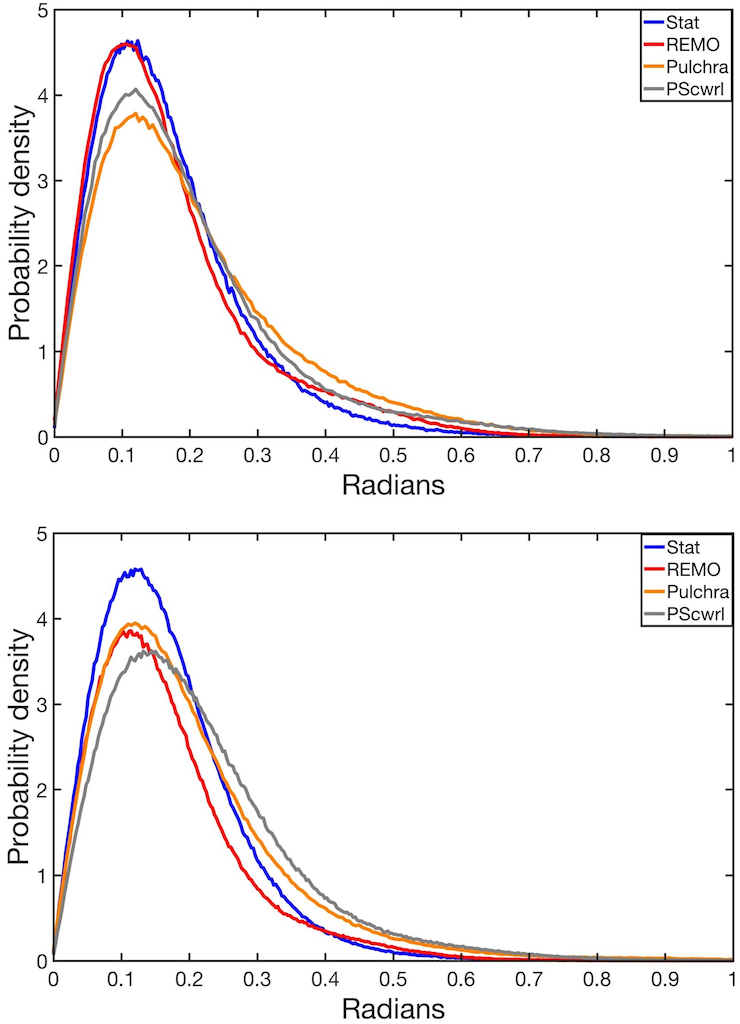}}
\caption {\small  {\it Color online:} 
Probability distribution for the individual angular deviations $\Theta^y_\beta[i;k]$ in the four reconstruction methods.Top: villin and Bottom: ww-domain. }
\label{fig-22}    
\end{figure}

In Figure \ref{fig-22} we have combined the probability distribution functions for the individual angles $\Theta^{\rm y}_{\beta}[i,k]$ in the case of C$\beta$,
for all the four reconstruction methods we consider. We observe very little difference between the four methods,  all distributions are strongly peaked near 
$\Theta \approx 0.15$ (rad). According to Figure \ref{fig-9} (bottom) the PDB average for the C$\alpha$-C$\beta$ bond length is around 1.54 \AA. Thus a $\Theta \approx 0.15$ (rad)
angular deviation corresponds to an average distance deviation of around 0.2 \AA ~ which is  
well within the the limits of the individual  B-factor fluctuation distances in Figure \ref{fig-8}.

%
%
%%%%%%%%%%%%
%%%%%%%%%%%%%%%
%%%%%%%%%%%%%%%%%%%%%%%%%%%%%%%%%%%%%%%%%%%%%
%%
%%
%%
%%
%%
%%                           FIGURE  22
%%
%%
%%
%%
%%
%% 
%%%%%%%%%%%%%%
%%%%%%%%%%%%%%
%
%
%
%%
%{
%\footnotesize
%
%
%
%
%%%%%%%%%%
%%%%%%%%%%%%%%
%%%%%%%%%%%%%%%

\subsubsection{RMS Probability densities for reconstructed chains}

\begin{figure}[h!]         
\centering            
  \resizebox{6.5cm}{!}{\includegraphics[]{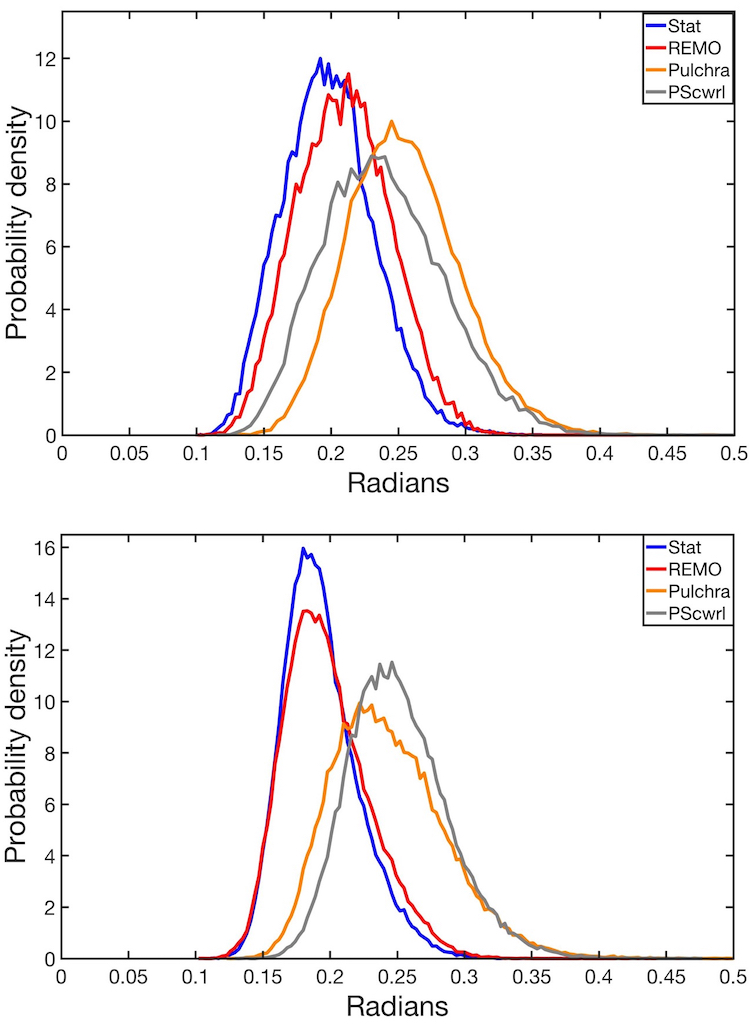}}
\caption {\small  {\it Color online:} 
Probability distribution for the angular RMS values (\ref{rmstheta}) 
in the four reconstruction methods. Top: villin and Bottom: ww-domain.}
\label{fig-23}    
\end{figure}

\begin{figure}[h!]         
\centering            
  \resizebox{6.5cm}{!}{\includegraphics[]{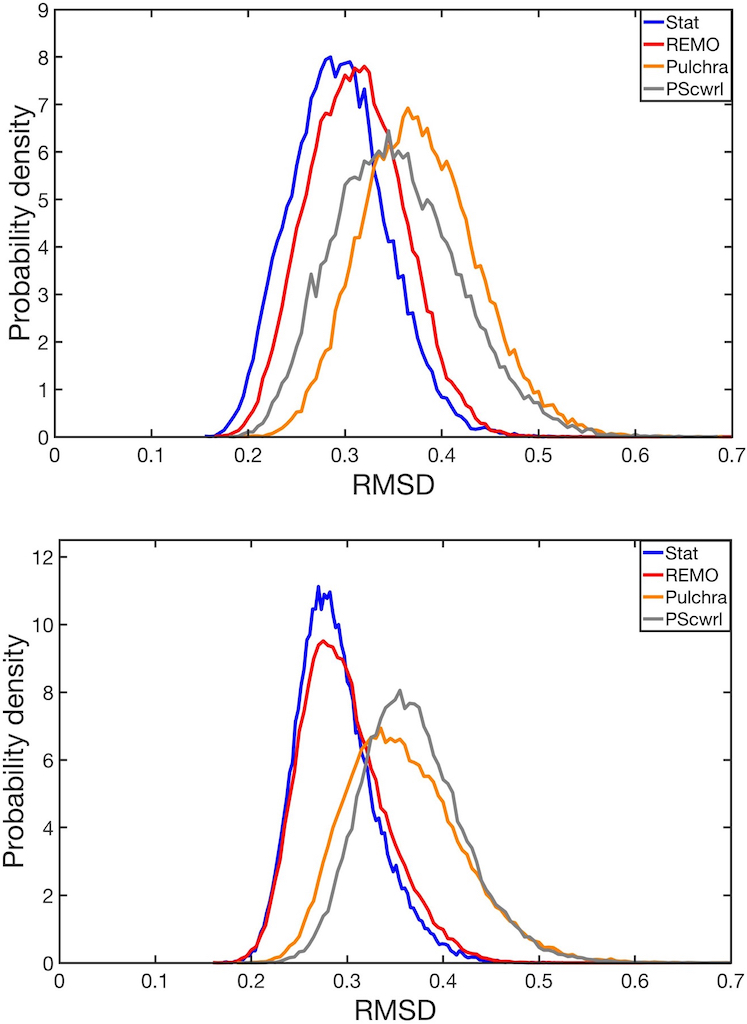}}
\caption {\small  {\it Color online:} 
Probability distribution for the RMS distances (\ref{RMSD}) 
in the four reconstruction methods. Top: villin and Bottom: ww-domain. } 
\label{fig-24}    
\end{figure}

Figures \ref{fig-23} show the C$\beta$ probability distributions of (\ref{rmstheta}) and Figures \ref{fig-24} show  the 
C$\beta$ probability distributions of (\ref{RMSD}), for the reconstructed chains. 
Generally speaking, all four methods are able to reconstruct the C$\beta$
positions with high  accuracy. The {\it Statistical Method} performs best but the difference to {\it Remo} is tiny.  For {\it Pulchra} the results are slightly worse:  Its combination
with {\it Scwrl4} performs a bit better than {\it Pulchra} alone in the case of villin, but in the case of ww-domain the results are opposite.   
However, the differences between all the four methods are quite small, and the RMS distances are 
all in line with the experimental B-factor fluctuation distances
shown in Figure \ref{fig-8}.

%
%
%%%%%%%%%%%%
%%%%%%%%%%%%%%%
%%%%%%%%%%%%%%%%%%%%%%%%%%%%%%%%%%%%%%%%%%%%%
%%
%%
%%
%%
%%
%%                           FIGURE  23
%%
%%
%%
%%
%%
%% 
%%%%%%%%%%%%%%
%%%%%%%%%%%%%%
%
%
%
%%
%{
%\footnotesize
%
%
%
%
%%%%%%%%%%
%%%%%%%%%%%%%%
%%%%%%%%%%%%%%%

%
%
%%%%%%%%%%%%
%%%%%%%%%%%%%%%
%%%%%%%%%%%%%%%%%%%%%%%%%%%%%%%%%%%%%%%%%%%%%
%%
%%
%%
%%
%%
%%                           FIGURE  24
%%
%%
%%
%%
%%
%% 
%%%%%%%%%%%%%%
%%%%%%%%%%%%%%
%
%
%
%%
%{
%\footnotesize
%
%
%
%
%%%%%%%%%%
%%%%%%%%%%%%%%
%%%%%%%%%%%%%%%

\section{Conclusions}

To comprehend protein dynamics is a prerequisite for the ability to understand how biologically active proteins function. However, 
despite the importance of protein dynamics, our knowledge  remains very limited. High quality experimental data on dynamical proteins
at near-physiological conditions is sparse and very difficult to come by, 
the primary source of information is theoretical considerations in combination with  
all atom molecular dynamics simulations; the latter are best exemplified by the very long {\it Anton} trajectories \cite{Lindorff-2011}
 
Here we have searched for systematics in the dynamics of proteins at  near-physiological conditions, by 
analyzing the {\it Anton} trajectories for villin and ww-domain.  We  have inquired to what extent can the dynamics of C$\alpha$ atoms 
determine that of the peptide plane O atoms and the side chain C$\beta$ atoms. For this we have compared the original simulation results with trajectories that we 
have reconstructed solely from the knowledge of the {\it Anton} C$\alpha$ trace. We have analyzed the results from  the reconstruction approaches  
{\it Pulchra}, {\it Remo}, {\it Scwrl} and a direct {\it Statistical Method} that we developed here. All these methods exploit crystallographic Protein Data Bank
structures which have been measured mostly at the very low temperatures of liquid nitrogen {\it i.e.}  below 77 K.  On the other hand, 
the {\it Anton} simulations have been performed at around 360K.  Thus we expect that besides  effects with a purely dynamical origin, 
there should be systematic differences that can be allocated to thermal fluctuations. 
Nevertheless, we  have found that the positions of both O and C$\beta$  atoms in the {\it Anton} 
trajectories can be determined with very high precision simply by using the knowledge of the static, crystallographic PDB structures; both dynamical and thermal effects 
are surprisingly small.  The results propose that the peptide plane and side chain dynamics is very strongly slaved to the C$\alpha$ atom motions, and 
subject to  only very small thermal fluctuation deviations.  

Our results can be explained in different ways: It would be truly remarkable if in a dynamical protein at near-physiological conditions, 
the O and C$\beta$ motions can indeed be determined, and with a very high precision, solely from a knowledge of the C$\alpha$ atom dynamics.  Such a strong
slaving to C$\alpha$ dynamics would be  a very strong impetus for the development of effective energy models for protein dynamics, 
in terms of reduced sets of coordinates at various levels of coarse graining. Alternatively, it can also be
that the force field CHARMM22$^\star$ that was used in the {\it Anton} simulations \cite{Lindorff-2011}, simply lacks the resiliency of actual proteins.  
In that case our results can shed light for ways to improve the accuracy of existing force field,  and help to determine more stringent standards for simulations. 
Indeed, we have observed the presence of very short-lived but systematic peptide plane flips along the {\it Anton} trajectories. These flips could be true physical effects that
are important to protein folding and dynamics. But they could as well be a consequence of  too harsh simulation obstructions, such as the use of too long elemental time steps and/or
exclusion of all fluctuations in the hydrogen covalent bond lengths. We have also observed, in the case of both {\it Pulchra} and {\it Statistical Method}, the presence of an apparent
two-state structure in the  O atom distributions, that seems to correlate with the distance between the dynamical structure and the natively folded state. In the case of
{\it Remo} no such two-stage structure is observed.

Quite unexpectedly to us, the purely PDB based {\it Statistical Method} appears to perform best in reconstructing the O and C$\beta$ atom positions. We suspect that this
is partly due to the choice of C$\alpha$ framing: The framings that are used in the case of {\it Pulchra} and {\it Remo} are mathematically correct, but  
might not account to the C$\alpha$ geometry as well as the Frenet framing does. It is possible that variants of {\it Pulchra} and {\it Remo} that are
based on Frenet framing, could bring about  even higher precision than any of the methods we have analysed. Thus,  our results 
should help the future development of increasingly precise reconstruction algorithms,  for a wide spectrum of refinement  and structure determination purposes.  
It  should also be of interest to extend the {\it Statistical Method} for the analysis of all other heavy atoms along a protein
structure, possibly following \cite{Peng-2014}.  

Finally, we note that the visual analysis methodology that we have developed is very versatile. It can be applied to analyze protein structure and dynamics, 
widely.

\section{Acknowledgements}

We thank N. Ilieva for communications and discussions, AJN also thanks A. Liwo for discussions. The work of AJN has been supported by the Qian Ren program at BIT, by a grant  2013-05288 and 2018-04411from Vetenskapsr\aa det, by Carl Trygger Stiftelse and
and by Henrik Granholms stiftelse.

\end{document}